\documentclass[journal]{IEEEtran}

\usepackage{lipsum}
\usepackage{graphicx}
\ifCLASSOPTIONcompsoc
    \usepackage[caption=false, font=normalsize, labelfont=sf, textfont=sf]{subfig}
\else
\usepackage[caption=false, font=footnotesize]{subfig}
\fi

\usepackage{graphicx}
\usepackage{amsfonts}
\usepackage{amsmath}
\usepackage{bm}
\usepackage{cite}
\usepackage{epstopdf}
\usepackage{booktabs}
\usepackage{lipsum}
\usepackage{flushend}
\usepackage{amssymb}
\usepackage{ntheorem}
\usepackage{algorithm}
\usepackage{algorithmic}
\usepackage{multirow}
\theorembodyfont{\rmfamily}
\newtheorem{lemma}{Lemma}
\newtheorem{fact}{Fact}
\newtheorem*{proof}{Proof}
\newtheorem*{remark}{Remarks}

\usepackage{xr}
\begin{document}

\title{DOA Estimation for Transmit Beamspace MIMO Radar via Tensor Decomposition with Vandermonde Factor Matrix}

\author{Feng Xu, \IEEEmembership{Student Member, IEEE}, Matthew W. Morency, \IEEEmembership{Student Member, IEEE}, and Sergiy A. Vorobyov, \IEEEmembership{Fellow, IEEE}
\thanks{This work was supported in part by the Academy of Finland under Grant 319822, in part by Huawei, and in part by the China Scholarship Council. This work was conducted while Feng Xu was a visiting doctoral student with the Department of Signal Processing and Acoustics, Aalto University. (\textit{Corresponding author: Sergiy A. Vorobyov.})}
\thanks{Feng Xu is with  the School of Information and Electronics, Beijing Institute of Technology, Beijing 100081, China, and also with the Department of Signal Processing and Acoustics, Aalto University, Espoo 02150, Finland. (e-mail: fengxu@bit.edu.cn, feng.xu@aalto.fi).}
\thanks{Matthew W. Morency is with the Dept. Microelectronics, School of Electrical Engineering, Mathematics, and Computer Science, Delft University of Technology, Mekelweg 4, 2628 CD Delft, The Netherlands. (e-mail: M.W.Morency@tudelft.nl).}
\thanks{Sergiy A. Vorobyov is with the Department of Signal Processing and Acoustics, Aalto University, Espoo 02150, Finland. (e-mail: svor@ieee.org).}
}
\maketitle

\begin{abstract}
We address the problem of tensor decomposition in application to direction-of-arrival (DOA) estimation for transmit beamspace (TB) multiple-input multiple-output (MIMO) radar. A general 4-order tensor model that enables computationally efficient DOA estimation is designed. Whereas other tensor decomposition-based methods treat all factor matrices as arbitrary, the essence of the proposed DOA estimation method is to fully exploit the Vandermonde structure of the factor matrices to take advantage of the shift-invariance between and within different subarrays. Specifically, the received signal of TB MIMO radar is expressed as a 4-order tensor. Depending on the target Doppler shifts, the constructed tensor is reshaped into two distinct 3-order tensors. A computationally efficient tensor decomposition method is proposed to decompose the Vandermonde factor matrices. The generators of the Vandermonde factor matrices are computed to estimate the phase rotations between subarrays, which can be utilized as a look-up table for finding target DOA. It is further shown that our proposed method can be used in a more general scenario where the subarray structures can be arbitrary but identical. The proposed DOA estimation method requires no prior information about the tensor rank and is guaranteed to achieve precise decomposition result. Simulation results illustrate the performance improvement of the proposed DOA estimation method as compared to conventional DOA estimation techniques for TB MIMO Radar.
\end{abstract}


\section{Introduction}\label{1}
\IEEEPARstart{T}{he} development of multiple-input multiple-output (MIMO) radar has been the focus of intensive research \cite{6,5,8,7,9} over the last decade, and has opened new opportunities in target detection and parameter estimation. Many works have been reported in the literature showing the applications of MIMO radar with widely separated antennas \cite{6} or collocated antennas \cite{5}. Among these applications, direction-of-arrival (DOA) estimation \cite{8,12,15,16,18,19,26,21,23} is one of the most fundamental research topics. In this paper, we mainly focus on the DOA estimation problem for MIMO radar with collocated antennas.

By ensuring that the transmitted waveforms are orthogonal \cite{11}, MIMO radar enables increasing the system's degree of freedom (DoF), improving the spatial resolution and enhancing the parameter identifiability. The essence behinds these advantages is the construction of a virtual array (VA), which can be regarded as a new array with larger aperture and more elements \cite{7,9}. However, the omnidirectional transmit beampattern in MIMO radar, resulting from the orthogonal waveforms, deteriorates the parameter estimation performance since most of the emitted energy is wasted as compared to its phased-array counterpart. To tackle this problem, the transmit beamspace (TB) technique has been introduced \cite{8,12,13}. In TB MIMO radar, the transmitted energy can be focused on a fixed region \cite{8,12} by using a number of linear combinations of the transmitted waveforms via a TB matrix. This benefit becomes more evident when the number of elements in MIMO radar is large \cite{13}. Specifically, at some number of waveforms, the gain from using more waveforms begins to degrade the estimation performance. The trade-off between waveform diversity and spatial diversity implies that the performance of DOA estimation in TB MIMO radar can be further improved with a carefully designed TB matrix.

Meanwhile, many algorithms for DOA estimation in MIMO radar have been proposed. These algorithms can be summarized in two categories, signal covariance matrix-based algorithms \cite{15,16,18,8,12,19} and signal tensor decomposition-based algorithms \cite{21,26,23,27,29,35,40,49,52}. For example, the estimation of target spatial angles can be conducted by multiple signal classification (MUSIC). The generalization of MUSIC to a planar array requires a 2-dimension (2-D) spectrum searching \cite{15}, and thus suffers from high computational complexity. By exploiting the rotational invariance property (RIP) of the signal subspace, estimation of signal parameters via rotational invariance technique (ESPRIT) \cite{8,12,16} can be applied to estimate the target angles without a spectrum searching. The RIP can be enforced in many ways, e.g., uniformly spaced antennas \cite{16} and the design of TB matrix \cite{8,12}. To further reduce the computational complexity and increase the number of snapshots, unitary-ESPRIT (U-ESPRIT) has been proposed \cite{19}. Some algorithms like propagator method (PM) have been studied \cite{18} to avoid the singular value decomposition (SVD) of the signal covariance matrix. The aforementioned DOA estimation algorithms are mostly conducted on a per-pulse basis to update the result from pulse to pulse. They ignore the multi-linear structure of the received signal in MIMO radar and, therefore, lead to poor performance in low signal-to-noise ratio (SNR) region.

The second category, signal tensor decomposition-based algorithms, has been proposed to address the problem of poor performance in low SNR. In particular, a 3-order tensor is introduced to store the whole received signal for MIMO radar in a single coherent processing interval (CPI). Methods like high-order SVD (HOSVD) \cite{23,33} and parallel factor  (PARAFAC) analysis \cite{21,26} can be applied to decompose the factor matrices. The DOA estimation can be conducted by exploiting the factor matrix with the target angular information. For example, the widely used alternating least square (ALS) algorithm is a common way of computing the approximate low-rank factors of a tensor. These factor matrices can be used to locate multiple targets simultaneously \cite{21,27}. Although the application of the conventional ALS algorithm improves the DOA estimation performance for MIMO radar, it usually requires the tensor rank as prior information, and the computational complexity can be extremely high as the convergence is unstable.

Nevertheless, conventional tensor decomposition methods are developed for tensors with arbitrary factor matrices. In array signal processing, special matrix structure like Toeplitz, Hankel, Vandermonde and columnwise orthonormal \cite{28,29} may exist in factor matrix when tensor model is applied to collect the received signal. The Vandermonde structure, as the most common one, can be generated from the application of carrier frequency offset, e.g., frequency diversity array (FDA) \cite{30} and orthogonal frequency-division multiplexing (OFDM) waveform \cite{55}, or uniformly spaced antennas, e.g., uniform linear array (ULA) and uniform rectangular array (URA). While conventional tensor decomposition methods are usually designed for tensors with arbitrary factor matrices, the decomposition of a tensor with structured factor matrices deserves further study as the structured factor matrix may point to a novel decomposition method and better uniqueness conditions. This is called {\it constrained tensor decomposition} \cite{28,29}. Moreover, transmit array interpolation is introduced for MIMO radar with arbitrary array structure \cite{23}. By solving the minimax optimization problem regarding interpolation matrix design, the original transmit array is mapped to a virtual array with desired structure. The DOA estimation bias caused by interpolation errors has also been analyzed in \cite{23}. However, the interpolation technique deteriorates the parameter identifiability, which makes it inappropriate for TB MIMO radar with arbitrary but identical subarrays.

In this paper, we consider the problem of tensor decomposition in application to DOA estimation for TB MIMO radar with multiple transmit subarrays.\footnote{Some preliminary ideas that have been extended and developed to this paper we published in \cite{31,53}.} A general 4-order tensor model that enables computationally efficient DOA estimation is designed. Whereas other tensor decomposition-based methods treat all factor matrices as arbitrary, the proposed DOA estimation method fully exploits the Vandermonde structure of the factor matrix to take advantage of the shift-invariance between and within different subarrays. In particular, the received signal of TB MIMO radar is expressed as a 4-order tensor. Depending on the target Doppler shifts, the constructed tensor is reshaped into two distinct 3-order tensors. A computationally efficient tensor decomposition method, which can be conducted via linear algebra with no iterations, is proposed to decompose the factor matrices of the reshaped tensors. Then, the Vandermonde structure of the factor matrices is utilized to estimate the phase rotations between transmit subarrays, which can be applied as a look-up table for finding target DOA. It is further shown that our proposed method can be used in a more general scenario where the subarray configurations are arbitrary but identical. By exploiting the shift-invariance, the proposed method improves the DOA estimation performance over conventional methods, and it has no requirement of prior information about the tensor rank. Simulation results verify that the proposed DOA estimation method has better accuracy and higher resolution.


The rest of this paper is organized as follows. Some algebra preliminaries about tensors and matrices are introduced at the end of Section~\ref{1}. A 4-order tensor model for TB MIMO radar with uniformly spaced subarrays is designed in Section~\ref{sig}. In Section~\ref{3}, the proposed tensor model is reshaped properly to achieve the uniqueness condition of tensor decomposition. The DOA estimation is conducted by exploiting the shift-invariance between and within different subarrays. The parameter identifiability is also analysed. Section~\ref{4} generalizes the proposed DOA estimation method to TB MIMO radar with non-uniformly spaced subarrays, where multiple scales of shift-invariances can be found. Section~\ref{5} performs the simulation examples while the conclusions are drawn in Section~\ref{6}.

\textsl{Notation}: Scalars, vectors, matrices and tensors are denoted by lower-case, boldface lower-case, boldface uppercase, and calligraphic letters, e.g., $y$, $\bf y$, $\bf Y$, and $\cal Y$, respectively. The transposition, Hermitian transposition, inversion, pseudo-inversion, Hadamard product, outer product, Kronecker product and Khatri-Rao (KR) product operations are denoted by ${\left( \cdot  \right)^T},{\left(  \cdot  \right)^H},{\left(  \cdot  \right)^{ - 1}}, {\left( \cdot  \right)^{\dag}}, * , \circ ,\otimes$, and $\odot$, respectively, while $vec\left(  \cdot  \right)$ stands for the operator which stacks the elements of a matrix/tensor one by one to a column vector. The notation $diag({\bf{y}})$ represents a diagonal matrix with its elements being the elements of ${\bf{y}}$, while $\left\| {\bf{Y}} \right\|_F$ and $\left\| {\bf{Y}} \right\|$ are the Frobenius norm and Euclidean norm of ${\bf{Y}}$, respectively. Moreover, ${{\bf{1}}_{M\times N}}$ and ${{\bf{0}}_{M\times N}}$ denote an all-one matrix of dimension $M \times N$ and an all-zero matrix of size $M \times N$, respectively, and ${{\bf{I}}_{M}}$ stands for the identity matrix of size $M \times M$. For ${\bf B} \in {{\mathbb{C}}^{M \times N}}$, the $n$-th column vector and $(m,n)$-th element are denoted by ${\bf b}_n$ and $B_{mn}$, respectively, while the $m$-th element of ${\bf b} \in {{\mathbb{C}^{M \times 1}}}$ is given by $b(m)$. The estimates of $\bf B$ and $\bf b$ are given by $\bf \hat B$ and $\bf \hat b$, while the rank and Kruskal-rank of ${\bf B}$ are denoted by $r({\bf B})$ and $k_{\bf B}$, respectively. To express two submatrices of $\bf B$ without the first and last row vector, ${\bf \underline B}$ and ${\bf \overline B}$ are applied. If the general form, $\bf B$ can be written as ${\bf B} \triangleq  [{\bm \beta}_1,{\bm \beta}_2,\cdots,{\bm \beta}_N]$, where ${\bm \beta}_n \triangleq  [1,z_n,z_n^2,\cdots,z_n^{M-1}]^T$, i.e., $\bf B$ is a Vandermonde matrix, and ${\bf z} \triangleq [z_1,z_2, \cdots, z_N]^T \in {{\mathbb{C}}^{N \times 1}} $ is the vector of generators. When each element is unique, ${\bf z}$ is considered to be distinct.

\subsection{Algebra Preliminaries for Tensors and Matrices}\label{2}
For an $N$-th order tensor ${{\cal Y} \in {{\mathbb{C}}^{{I_1} \times {I_2}\times \cdots \times {I_N}}}}$, the following facts are introduced \cite{27,34}.
\begin{fact}(PARAFAC decomposition):
The PARAFAC decomposition of an $N$-th order tensor is a linear combination of the minimum number of rank-one tensors, given by
\begin{equation}
\begin{aligned}
{\cal Y}  =  \sum\limits_{l = 1}^L {{{\bm{\alpha}}_l^{(1)}} \circ {{\bm{\alpha}}_l^{(2)}} \circ \cdots \circ {{\bm{\alpha}}_l^{(N)}}}\triangleq  [[{\bf A}^{(1)}, {\bf A}^{(2)},\cdots, {\bf A}^{(N)}]]
\end{aligned}
\end{equation}\label{PARAFAC}
\end{fact}
where ${{\bm{\alpha}}_l^{(n)}}$ is the $l$-th column of ${\bf{A}}^{(n)}$ with ${\bf{A}}^{(n)}$ being the $n$-th factor matrix of size $I_n \times L$, and $L$ is the tensor rank.

\begin{fact} (Uniqueness of PARAFAC decomposition):\label{krank}
The PARAFAC decomposition is unique if all potential factor matrices satisfying \eqref{PARAFAC} also match with
\begin{equation}
{{\bf{\tilde A}}^{(n)}} = {{\bf{A}}^{(n)}}{{\bf{\Pi }}^{(n)}}{{\bf{\Delta }}^{(n)}}
\end{equation}
where ${{\bf{\Pi }}^{(n)}}$ is a permutation matrix and ${{\bf{\Delta }}^{(n)}}$ is a diagonal matrix. The product of ${{\bf{\Delta }}^{(n)}},n = 1,2,\cdots,N$ is an $L \times L$ identity matrix. Usually, the generic uniqueness condition is given by \cite{34}: $\sum\limits_{n = 1}^N {{k_{{{\bf{A}}^{(n)}}}}}  \ge 2L + (N - 1)$.
\end{fact}

\begin{fact}(Mode-$n$ unfolding of tensor):\label{tensordef}
The mode-$n$ unfolding of a tensor ${{\cal Y} \in {{\mathbb{C}}^{{I_1} \times {I_2}\times \cdots \times {I_N}}}}$ is denoted by ${\bf Y}_{(n)}$, which is a matrix of size ${{I_1}\cdots{I_{n-1}}{I_{n+1}}\cdots{I_N}}\times {I_n}$
\begin{equation}
{\bf Y}_{(n)} = \left( {{{\bf{A}}^{(1)}} \cdots \odot {{\bf{A}}^{(n - 1)}} \odot {{\bf{A}}^{(n + 1)}} \cdots \odot{{\bf{A}}^{(N)}}} \right)\left({{{{\bf{A}}^{(n)}}}}\right)^T.
\end{equation}
\end{fact}

\begin{fact}\label{reshape} (Tensor reshape):
The reshape operator for an $N$-th order tensor ${{\cal Y} \in {{\mathbb{C}}^{{I_1} \times {I_2}\times \cdots \times {I_N}}}}$ returns a new $M$-th order tensor ${{\cal X} \in {{\mathbb{C}}^{{J_1} \times {J_2}\times \cdots \times {J_M}}}}$ ($M \le N$) with $\prod\limits_{n = 1}^N {{I_n}}  = \prod\limits_{m = 1}^M {{J_m}}$ and $vec({\cal Y}) = vec({\cal X})$, e.g., if ${J_m} = {I_m}, m = 1,2,\cdots,M-1$ and ${J_M} = \prod\limits_{n = M}^N {{I_n}}$, the mode-$M$ unfolding of reshaped $\cal X$ is
\begin{equation}
{{\bf{X}}_{(m)}} = \left( {{{\bf{A}}^{(1)}} \odot \cdots \odot {{\bf{A}}^{(M-1)}}} \right){\left( {{{\bf{A}}^{(M)}} \odot \cdots \odot {{\bf{A}}^{(N)}}} \right)^T}.
\end{equation}
\end{fact}

\begin{lemma}:\label{111}
For a 3-order tensor ${\cal Y} \triangleq [[{\bf A}^{(1)},{\bf A}^{(2)},{\bf A}^{(3)}]]$, where ${\bf A}^{(1)}$ is a Vandermonde matrix or the KR product of two Vandermonde matrices. The decomposition of $\cal Y$ is generically unique if the generators of ${\bf A}^{(1)}$ are distinct and ${\bf A}^{(3)}$ is column full rank.
\begin{proof}
It is purely technical and is given in supplemental material as Appendix~\ref{222}.
\end{proof}
\end{lemma}

\begin{lemma}: The following equalities hold true
\begin{equation}
\begin{aligned}
&{\bf A}{\bf B} = {\bf A} \odot {\bf b}^T = {\bf b}^T \odot {\bf A}\\
& {\bf A} \odot {\bf b}^T\odot {\bf C} = {\bf b}^T \odot {\bf A}\odot {\bf C} = {\bf A} \odot {\bf C}\odot {\bf b}^T\\
& ({\bf A} \odot {\bf B})\odot {\bf C} = {\bf A} \odot ({\bf B}\odot {\bf C})\\
& vec\left( {{\bf{ABD}}} \right) = \left( {{{\bf{D}}^T} \odot {\bf{A}}} \right){\bf{b}}\\
& \left( {{\bf{A}} \otimes {\bf{C}}} \right)\left( {{\bf{D}} \otimes {\bf{E}}} \right) = \left( {{\bf{AD}}} \right) \otimes \left( {{\bf{CE}}} \right)
\end{aligned}
\end{equation}
where ${\bf{A}} \in {\mathbb{C}^{M \times N}}$, ${\bf{C}} \in {\mathbb{C}^{Q \times N}}$, ${\bf{D}} \in {\mathbb{C}^{N \times P}}$, ${\bf{E}} \in {\mathbb{C}^{N \times L}}$ and ${\bf{B}} = diag({\bf{b}}) \in {\mathbb{C}^{N \times N}}$.
\end{lemma}

\section{TB MIMO Radar Tensor model}\label{sig}
\subsection{TB MIMO Radar with Linear Array}
Consider a collocated MIMO radar with $M$ transmit elements and $N$ receive elements. The transmit array is a ULA with its elements spaced at half the working wavelength away from each other. The receive elements are randomly placed within a fixed aperture. Assuming $S$ subarrays are uniformly spaced at the transmit side and also assuming that each subarray contains $M_0$ elements, the indices of first elements in those subarrays are denoted by $m_s,s = 1,2,\cdots,S$. Without loss of generality, $m_s$ rises uniformly. The steering vectors of the entire transmit array and the first transmit subarray at direction $\theta$ can be given by ${\bm{\alpha }}(\theta ) \triangleq {\left[ {1,{e^{ - j\pi\sin \theta }},\cdots,{e^{ - j(M-1)\pi\sin \theta }}} \right]^T}$ and ${{\bm{\alpha }}_0}(\theta) \triangleq {\left[ {1,{e^{ - j\pi\sin \theta }},\cdots,{e^{ - j(M_0-1)\pi\sin \theta }}} \right]^T}$, respectively. The steering vector of the receive array can be written as ${\bm{\beta }}(\theta ) \triangleq {\left[ 1,{{e^{ - j\frac{{2\pi }}{\lambda }{x_2}\sin \theta}},\cdots,{e^{ - j\frac{{2\pi }}{\lambda }{x_{N}}\sin \theta }}} \right]^T}$, where $\left\{ {\left. x_{n} \right|{\rm{0}} < {x_{n}} \le {D}, n = 1,\cdots,N} \right\}$ and $D$ is the aperture of the receive array.

Accordingly, the transmit and receive steering matrices for $L$ targets in $\left\{ {{\theta _{l}}} \right\}_{l = 1}^L$  can be denoted by ${\bf{A}} \triangleq \left[ {{\bm{\alpha }}({\theta _1}),{\bm{\alpha }}({\theta _2}),\cdots,{\bm{\alpha }}({\theta _L})} \right]$ and ${\bf{B}} \triangleq \left[ {{\bm{\beta }}({\theta _1}),{\bm{\beta }}({\theta _2}),\cdots,{\bm{\beta }}({\theta _L})} \right]$, respectively, while the steering matrix for the first transmit subarray can be given as ${\bf{A}}_0 \triangleq \left[ {{\bm{\alpha }}_0({\theta _1}),{\bm{\alpha }}_0({\theta _2}),\cdots,{\bm{\alpha }}_0({\theta _L})} \right]$. Note that ${\bf{A}}_0$ can also be regarded as the submatrix of ${\bf{A}}$ with first $M_0$ rows.

In conventional MIMO radar, the received signal at the output of the receive array after matched-filtering in matrix form can be modelled as \cite{21}: ${\bf{Y}} = {\bf{B\Sigma}}{\bf A}^T + {\bf N}$, where ${\bf {\Sigma}} = diag({\bm \sigma})$, ${\bm{\sigma}} \triangleq \left[ {\sigma _1^2,\sigma _2^2,\cdots,\sigma _L^2} \right]^T$ represents the vector of target radar cross section (RCS) fading coefficients obeying Swerling I model, and ${\bf  N}$ is the noise residue of size $N\times M$. When the TB technique is introduced \cite{8,12}, the received signal model after matched-filtering of $K$ orthogonal waveforms ($K \le M$) can be generalized as ${\bf{Y}} = {\bf{B\Sigma}}({{\bf W}^ H\bf A})^T + {\bf  N}$, where ${\bf{W}} \triangleq {\left[ {{{\bf{w}}_1},{{\bf{w}}_2},\cdots,{{\bf{w}}_K}} \right]_{M \times K}}$ denotes the TB matrix.

Hence, the received signal for the $s$-th transmit subarray, $s = 1,2,\cdots, S$, and the whole receive array can be written as
\begin{equation}
{\bf{Y}}_s = {\bf{B\Sigma}}({{\bf W}_s^ H {\bf A}_s})^T + {\bf N}_s \label{matrix}
\end{equation}
where ${\bf W}_s$ and ${\bf A}_s$ represent the TB matrix and steering matrix for the $s$-th transmit subarray, respectively, and ${\bf N}_s$ is the noise residue of size $N\times {M_0}$. Assume that the TB matrix for each subarray is identical, denoted by ${\bf W}_0 \in {\mathbb{C}^{{M_0} \times K}}$. Note that since $m_s$ rises uniformly, the steering matrix for the $s$-th transmit subarray can also be expressed as ${\bf A}_s = {\bf A}_0{\bf \Gamma}_s$, where ${\bf \Gamma}_s = diag({{\bf{k}}_s})$ and $ {{\bf{k}}_s} \triangleq \left[ {{e^{ - j\pi ({m_s} - 1)\sin {\theta _1}}},\cdots,{e^{ - j\pi ({m_s} - 1)\sin {\theta _L}}}} \right]^T$. Substituting this relationship into \eqref{matrix} and vectorizing it, we have ${{\bf{y}}_{s}^{}}  = \left[ {\left( {{\bf{W}}_0^H{{\bf{A}}_0}} \right) \odot {\bf{B}}} \right]{\bf \Gamma}_s{{\bm \sigma}} + {{\bf{n}}_{s}^{}}$, where ${\bf n}_s$ is the vectorized noise residue.

Considering the Doppler effect, the received signal during $q$-th pulse in a single CPI, $q = 1,2,\cdots,Q$, can be written as
\begin{equation}
{{\bf{y}}_{s}^{(q)}}  = \left[ {\left( {{\bf{W}}_0^H{{\bf{A}}_0}} \right) \odot {\bf{B}}} \right]{\bf \Gamma}_s{{\bf c}_q} + {{\bf{n}}_{s}^{(q)}}\label{vector_subarray}
\end{equation}
where ${{\bf{c}}_{q}} = {\bm{\sigma}} * {{\bf{\bar c}}_{{q}}}$, ${{\bf{\bar c}}_{{q}}} \triangleq \left[ {{e^{i2\pi {f_1}{q}T}},{e^{i2\pi {f_2}{q}T}},\cdots,{e^{i2\pi {f_L}{q}T}}} \right]^T$, $f_l$ denotes the Doppler shift, $T$ is the radar pulse duration, and ${{\bf{n}}_{s}^{(q)}}$ is the vectorized noise residue. Concatenate the received signal of $S$ subarrays in $q$-th pulse, i.e., ${{\bf{Y}}^{(q)}} \triangleq \left[ {{{\bf{y}}_{1}^{(q)}},{{\bf{y}}_{2}^{(q)}},\cdots,{{\bf{y}}_{S}^{(q)}}} \right]_{KN \times S}$. The compact form can be written as
\begin{equation}
{{\bf{Y}}^{(q)}} = \left[ {\left( {{\bf{W}}_0^H{{\bf{A}}_0}} \right) \odot {\bf{B}}} \right]{\left[ {{{\bf{c}}_q^T} \odot {\bf{K}}} \right]^T} + {{\bf{N}}^{(q)}}\label{Subarray}
\end{equation}
where ${\bf{K}} \triangleq {\left[ {{\bf{k}}_1,{\bf{k}}_2,\cdots,{\bf{k}}_S} \right]^T}_{S \times L}$ and ${{\bf{N}}^{(q)}} \triangleq \left[ {{{\bf{n}}_{1}^{(q)}},{{\bf{n}}_{2}^{(q)}},\cdots,{{\bf{n}}_{S}^{(q)}}} \right]$. Note that the $l$-th column of $\bf K$ represents the phase rotations of $l$-th target for $S$ subarrays. $\bf K$ can be named as {\it transmit subarray steering matrix}.

Vectorizing \eqref{Subarray}, the $KNS \times 1$ vector can be given as
\begin{equation}
\begin{aligned}
{{\bf{z}}_q} & = \left\{ {\left[ {{{\bf{c}}_q^T} \odot {\bf{K}}} \right] \odot \left[ {\left( {{\bf{W}}_0^H{{\bf{A}}_0}} \right) \odot {\bf{B}}} \right]} \right\}{{\bf{1}}_{L \times 1}}+{\bf r}_q\\
& =  \left[ {{\bf{K}} \odot \left( {{\bf{W}}_0^H{{\bf{A}}_0}} \right) \odot {\bf{B}}} \right]{\bf{c}}_q + {{\bf{r}}_q}
\end{aligned}\label{ULAz}
\end{equation}
where ${{\bf{r}}_q}$ is the vectorized noise residue of ${{\bf{N}}^{(q)}}$. Then, concatenate the received signal of $Q$ pulses, i.e., ${{\bf{Z}}} \triangleq \left[ {{{\bf{z}}_{1}},{{\bf{z}}_{2}},\cdots,{{\bf{z}}_{Q}}} \right]_{KNS \times Q}$. The compact form can be formulated as
\begin{equation}
{{\bf{Z}}} = \left[ {{\bf{K}} \odot \left( {{\bf{W}}_0^H{{\bf{A}}_0}} \right) \odot {\bf{B}}} \right]{{\bf{C}}^T} + {{\bf{R}}}\label{unfolding}
\end{equation}
where ${{\bf{C}}} \triangleq {\left[ {{\bf{c}}_1,{\bf{c}}_2,\cdots,{\bf{c}}_{Q}} \right]^T}_{Q \times L}$ and ${{\bf{R}}} \triangleq \left[ {{{\bf{r}}_{1}},{{\bf{r}}_{2}},\cdots,{{\bf{r}}_{Q}}} \right]$. Similarly, $\bf C$ can be named as {\it Doppler steering matrix} since each column denotes the Doppler steering vector for one target (with additional RCS information). According to \textbf{Fact~\ref{tensordef}}, a 4-order tensor ${\cal Z} \in {\mathbb{C}^{{S} \times K \times N\times{Q}}}$ whose matricized version is ${{\bf{Z}}}$ in \eqref{unfolding} can be constructed. Denote ${\bf X} \triangleq {{\bf{W}}_0^H{{\bf{A}}_0}}$, then this tensor can be written as
\begin{equation}
{\cal Z} =  \sum\limits_{l = 1}^L {{{\bm{\kappa }}_l} \circ {{\bm{\chi}}_l} \circ {{\bm{\beta }}_l} \circ {{\bm{\gamma }}_l}}+{\cal R} \triangleq [[ {{\bf{K}},{\bf{X}},{\bf{B}},{\bf{C}}}
]] +{\cal R}\label{tensor1}
\end{equation}
where ${{{\bm{\kappa }}_l}, {{\bm{\chi}}_l}, {{\bm{\beta }}_l}, {{\bm{\gamma }}_l}}$ are the $l$-th columns of ${\bf{K}},{\bf{X}},{\bf{B}},{\bf{C}}$, respectively, $L$ is the tensor rank, and $\cal R$ is the noise tensor of the same size.

\subsection{TB MIMO Radar with Planar Array}
\begin{figure}
\centerline{\includegraphics[width=0.8 \columnwidth]{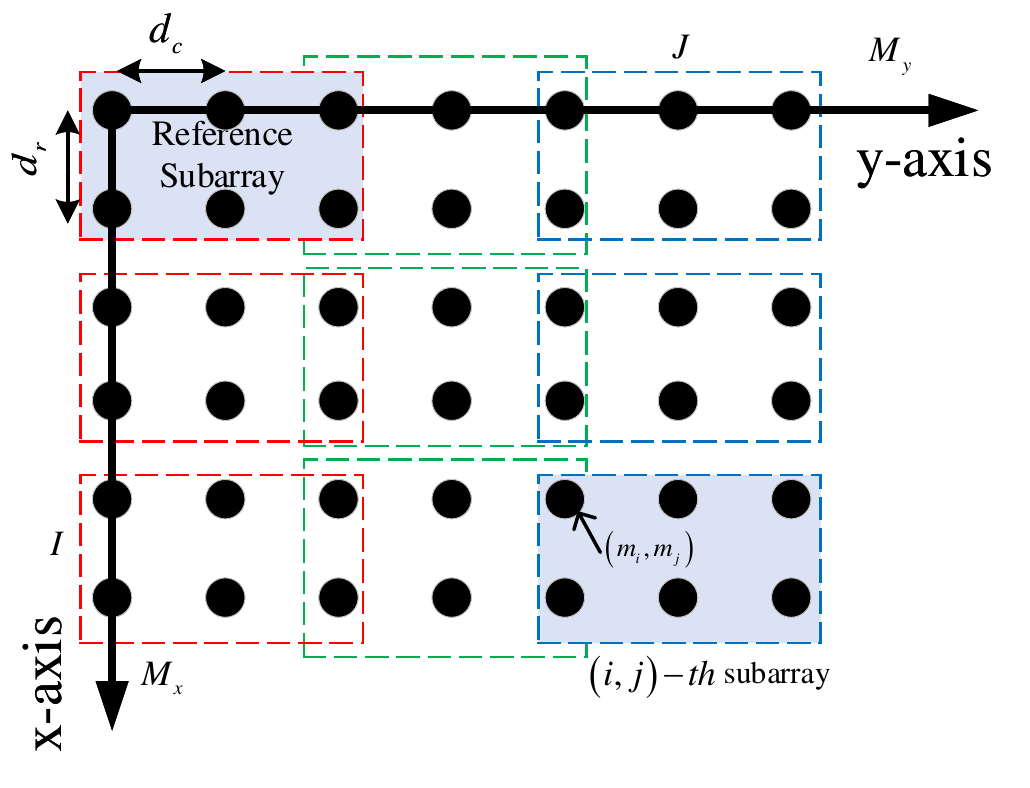}}
\caption{Transmit array configuration for TB MIMO radar with planar array.}\label{sys}
\end{figure}

Consider the planar array case, as shown in Fig.~\ref{sys}. A URA with $M = {M_x}\cdot{M_y}$ elements spaced at half the working wavelength in both directions and a planar array with $N$ randomly spaced elements are applied as the transmit and receive array, respectively. The transmit steering vector can be given as ${\bm{\alpha }}(\theta ,\varphi ) = {\bf{u}}(\theta ,\varphi ) \otimes {\bf{v}}(\theta ,\varphi )$, where ${\bf{u}}(\theta ,\varphi ) \triangleq {\left[ {1,{e^{ - j\pi u }},\cdots,{e^{ - j({M_y} - 1)\pi u}}} \right]^T}$, ${\bf{v}}(\theta ,\varphi ) \triangleq {\left[ {1,{e^{ - j\pi v}},\cdots,{e^{ - j({M_x} - 1)\pi v }}} \right]^T}$, $u \triangleq \sin \varphi \sin \theta$, $v \triangleq \sin \varphi \cos \theta $, and $(\theta ,\varphi )$ is the pair of azimuth and elevation of a target. The steering vector of the receive array can be written as ${\bm{\beta }}(\theta, \varphi ) \triangleq {\left[ {1,{e^{ - j\frac{{2\pi }}{\lambda }({x_{2}}v  + {y_{2}}u)}}, \cdots,{e^{ - j\frac{{2\pi }}{\lambda }({x_{N}}v  + {y_{N}}u)}}} \right]^T}$, where $\left\{ {\left. {({x_{n }},{y_{n }})} \right|{\rm{0}} < {x_{n}} \le {D_x},{\rm{0}} < {y_{n}} \le {D_y}} \right\}$ are the coordinates of the receive elements, and $D_x, D_y$ denote the apertures in two directions, respectively.

Accordingly, assume $S = I\cdot J$ transmit subarrays are uniformly spaced at the transmit side, which can be overlapped or not. Each of them contains $M_0 = {M_{x_0}}\cdot {M_{y_0}}$ elements. The first subarray is selected as the reference subarray. For $L$ targets in $\left\{\left( {{\theta _{l}}}, {\varphi_l} \right) \right\}_{l = 1}^L$, the transmit and receive steering matrices can be generalized as ${\bf{A}} \triangleq \left[ {{\bm{\alpha }}({\theta _1},{\varphi_1}),{\bm{\alpha }}({\theta _2},{\varphi_2}),\cdots,{\bm{\alpha }}({\theta _L},{\varphi_L})} \right]$ and ${\bf{B}} \triangleq \left[ {{\bm{\beta }}({\theta _1},{\varphi_1}),{\bm{\beta }}({\theta _2},{\varphi_2}),\cdots,{\bm{\beta }}({\theta _L},{\varphi_L})} \right]$, respectively. Note that the transmit array is a URA, thus, we have ${\bf{A}} = {{\bf{U}}} \odot {{\bf{V}}}$, where ${{\bf{U}}} \triangleq \left[ {{\bf{u}}({\theta _1},{\varphi _1}),{\bf{u}}({\theta _2},{\varphi _2}),\cdots,{\bf{u}}({\theta _L},{\varphi _L})} \right]$ and ${{\bf{V}}} \triangleq \left[ {{\bf{v}}({\theta _1},{\varphi _1}),{\bf{v}}({\theta _2},{\varphi _2}),\cdots,{\bf{v}}({\theta _L},{\varphi _L})} \right]$. Similarly, the steering vector of the reference transmit subarray can be written as ${\bm{\alpha }}_0(\theta ,\varphi ) = {\bf{u}}_0(\theta ,\varphi ) \otimes {\bf{v}}_0(\theta ,\varphi )$, where ${\bf{u}}_0(\theta ,\varphi )$ and ${\bf{v}}_0(\theta ,\varphi )$ contain the first $M_{y_0}$ and $M_{x_0}$ elements in ${\bf{u}}(\theta ,\varphi )$ and ${\bf{v}}(\theta ,\varphi )$, respectively. The steering matrix for the reference transmit subarray can be denoted by ${\bf{A}}_0 = {{\bf{U}}_{0}} \odot {{\bf{V}}_{0}}$, where ${{\bf{U}}_{0}}$ and ${{\bf{V}}_{0}}$ are the submatrices of ${{\bf{U}}}$ and ${{\bf{V}}}$ that consist of the first $M_{y_0}$ and $M_{x_0}$ rows, respectively.

For the $(i,j)$-th subarray (or equivalently, for the $s$-th transmit subarray where $s = (j-1)I+i$), the index of first element is denoted by $(m_{i},m_{j}),\ i = 1,2,\cdots,I,\ {j} = 1,2,\cdots, J$. Both $m_{i}$ and $m_{j}$ rise uniformly. The steering matrix for the $(i,j)$-th subarray can be given as ${\bf{A}}_{ij} = {{\bf{U}}_{j}} \odot {{\bf{V}}_{i}}$, where ${{\bf{U}}_{j}} = {{\bf{U}}_{0}}{{\bf \Gamma}_{j}}$, ${{\bf{V}}_{i}} = {{\bf{V}}_{0}}{{\bf \Gamma}_{i}}$, ${{\bf \Gamma}_{j}} = diag({\bf h}_{j}),{{\bf \Gamma}_{i}} = diag({\bf d}_{i})$, vectors ${\bf h}_{j} \triangleq {\left[{{e^{ - j\pi{(m_{j}-1)}u_1 }},\cdots,{e^{ - j\pi{(m_{j}-1)}u_L }}} \right]^T}$ and ${\bf d}_{i} \triangleq {\left[ {{e^{ - j\pi{(m_{i}-1)}v_1 }},\cdots,{e^{ - j\pi{(m_{i}-1)}v_L }}} \right]^T}$ indicate the phase rotations for $L$ targets in two directions, respectively.

Generalizing \eqref{vector_subarray}, the received signal after matched-filtering for the $s$-th transmit subarray and the whole receive array in $q$-th pulse can be written as
\begin{equation}
{{\bf{y}}^{(q)}_{s}}  = \left[ {\left( {{\bf{W}}_0^H{{\bf{A}}_0}} \right) \odot {\bf{B}}} \right]{{\bf \Gamma}_{i}}{{\bf \Gamma}_{j}}{{\bf c}_q} + {{\bf{n}}^{(q)}_{s}}.\label{z}
\end{equation}

Similarly, the concatenation of the received signal ${{\bf{y}}^{(q)}_{s}}$ for all $S$ subarrays in $q$-th pulse can be expressed as
\begin{equation}
{\bf Y}^{(q)} = \left[ {\left( {{\bf{W}}_0^H{{\bf{A}}_0}} \right) \odot {\bf{B}}} \right]{\left( {{{\bf{c}}_q^T} \odot {\bf{H }} \odot {\bf{\Delta}}} \right)^T} + {\bf N}^{(q)}\label{2Dunfolding}
\end{equation}
where ${\bf{H}} \triangleq {\left[ {{\bf{h}}_1,{\bf{h}}_2,\cdots,{\bf{h}}_{J}} \right]^T}_{{J} \times L}$ and ${\bf{\Delta}}\triangleq {\left[ {{\bf{d}}_1,{\bf{d}}_2,\cdots,{\bf{d}}_{I}} \right]^T}_{{I} \times L}$. Proof of \eqref{2Dunfolding} is purely technical and is given in supplemental material as Appendix~\ref{A}. Then ${{\bf{z}}_q} = vec({\bf Y}^{(q)})$ can be formulated as
\begin{equation}
{{\bf{z}}_q} = \left[ {{\bf{H}} \odot {\bf \Delta} \odot \left( {{\bf{W}}_0^H{{\bf{A}}_0}} \right) \odot {\bf{B}}} \right]{\bf{c}}_q + {{\bf{r}}_q}. \label{yq}
\end{equation}

After concatenating ${{\bf{z}}_q}$ in the same way as \eqref{unfolding}, the received signal of $Q$ pulses in the URA case can be written as
\begin{equation}
{{\bf{Z}}} = \left[ {{\bf{H}}\odot {\bf{\Delta}} \odot \left( {{\bf{W}}_0^H{{\bf{A}}_0}} \right) \odot {\bf{B}}} \right]{{\bf{C}}^T} + {{\bf{R}}}.\label{unfolding2}
\end{equation}

It is interesting that, \eqref{unfolding2} can be directly obtained from \eqref{unfolding} by replacing $\bf K$ with ${\bf H} \odot {\bf \Delta}$. Hence, ${\bf H} \odot {\bf \Delta}$ can be regarded as the {\it transmit subarray steering matrix} for URA. Using \textbf{Fact~\ref{tensordef}}, a 5-order tensor $\cal Z$ whose matricized version is ${{\bf{Z}}}$ in \eqref{unfolding2} can be constructed as
\begin{equation}
{\cal Z}  = \sum\limits_{l = 1}^L {{{\bm{\eta}}_l} \circ{{\bm{\delta }}_l} \circ {{\bm{\chi}}_l} \circ {{\bm{\beta }}_l} \circ {{\bm{\gamma }}_l}}+{\cal R}\triangleq [[{{\bf{H}},{\bf{\Delta}},{\bf{X}},{\bf{B}},{\bf{C}}}
 ]] +{\cal R} \label{tensor2}
\end{equation}
where ${{\bm{\eta}}_l}$ and ${{\bm{\delta}}_l}$ are the $l$-th columns of ${\bf{H}}$ and ${\bf{\Delta}}$, respectively.

Note that since all subarrays are uniformly spaced, ${\bf{K}}$, ${\bf{\Delta}}$ and ${\bf{H}}$ are Vandermonde matrices and their vectors of generators can be respectively denoted by
\begin{equation}
\begin{aligned}
&{{\bm{\omega}}} \triangleq {\left[ {{e^{ - j\pi {\Delta_m}\sin {\theta _1}}},\cdots,{e^{ - j\pi {\Delta_m}\sin {\theta _L}}}} \right]^T}\\
&{{\bm{\omega}}_x} \triangleq {\left[ {{e^{ - j\pi {\Delta_{m_x}}v_1}},\cdots,{e^{ - j\pi {\Delta_{m_x}}v_1}}} \right]^T}\\
&{{\bm{\omega}}_y} \triangleq {\left[ {{e^{ - j\pi {\Delta_{m_y}}u_1}},\cdots,{e^{ - j\pi {\Delta_{m_y}}u_L}}} \right]^T}
\end{aligned}\label{generate}
\end{equation}
where the step sizes ${\Delta}_m = m_{s+1}-m_s$, ${\Delta_{m_x}} =  m_{{i}+1}-m_{i}$, and ${\Delta_{m_y}} =  m_{{j}+1}-m_{j}$. We assume ${{\bm{\omega}}}$, ${{\bm{\omega}}}_x$ and ${{\bm{\omega}}}_y$ are distinct, which means that multiple targets are spatially distinct.

\section{DOA Estimation via Tensor Decomposition with Vandermonde Factor Matrix}\label{3}
We have shown that the received signal of TB MIMO radar with transmit subarrays can be formulated as a high-order tensor. It is useful to point out that \eqref{tensor1} and \eqref{tensor2} are identical if the idea of tensor reshape is applied and ${\bf K}$ is replaced by ${\bf H} \odot {\bf \Delta}$. Hence, a general 4-order tensor model can be used to express the received signal for TB MIMO radar with uniformly spaced subarrays, given by
\begin{equation}
{\cal Z} \triangleq \left[ [{{\bf{G}},{\bf{X}},{\bf{B}},{\bf{C}}}
 \right]]+{\cal R}\label{general}
\end{equation}
where ${\bf{G}}\in {\mathbb{C}^{{S} \times {L}}}$ is the {\it transmit subarray steering matrix}. Essentially, $\bf G$ can be interpreted as the result of element-wise spatial smoothing between the transmit elements. A new dimension is extended to express the phase rotations between transmit subarrays in the tensor model, which matches with the derivations in \eqref{tensor1} and \eqref{tensor2}.

The tensor decomposition of  ${\cal Z}$ can be regarded as the constrained tensor decomposition, since one of the factor matrices is structured by the regular array configuration. Generally, the ALS algorithm can be applied to decompose such a tensor. However, the convergence of the ALS algorithm heavily relies on the determination of tensor rank, which is an NP-hard problem. The Vandermonde structure of the factor matrix is ignored. The number of iterations in ALS algorithm is also uncertain, which may lead to high computational complexity. In the literature \cite{28,29,35}, the uniqueness condition of the tensor decomposition with special-structured factor matrices, e.g., Toeplitz, Hankel, Vandermonde and column-wise orthonormal, has been investigated. The structured factor matrix may change the uniqueness condition and, therefore, point to some new tensor decomposition methods.

In this section, we mainly focus on the tensor decomposition with Vandermonde factor matrix in application to DOA estimation for TB MIMO radar with uniformly spaced subarrays. A computationally efficient DOA estimation method is proposed and we discuss the application of the proposed method for both linear and planar arrays.

To begin with, a 3-order tensor ${\cal F} \triangleq [[ {{\bf{G}},({\bf{X}} \odot {\bf{B}}), {\bf{C}})}]]$ can be reshaped from \eqref{general} (see \textbf{Fact~\ref{reshape}}), whose mode-3 unfolding is ${\bf F}_{(3)} = ({\bf G} \odot {\bf{X}} \odot {\bf{B}}){\bf C}^T$. Considering only the $q$-th pulse, ${\bf F}_{(3)}$ is generically identical to that in \eqref{ULAz} for linear array or \eqref{yq} for planar array. In other words, the signal covariance matrix-based DOA estimation methods like MUSIC and \-ESPRIT can be conducted by using ${\bf R} = 1/Q{\bf F}_{(3)}{\bf F}^H_{(3)}$ as the signal covariance matrix. Meanwhile, note that ${\bf{G}}$ is either a Vandermonde matrix or the KR product of a pair of Vandermonde matrices. Thus, \textbf{Lemma~\ref{111}} can be applied to conduct a tensor decomposition-based DOA estimation if the second precondition is satisfied, i.e., $r({\bf C}) = L$.

Take a ULA for example, let ${\bf G} = {\bf K}$. The SVD of  ${{\bf{F}}_{(3)}}$ is denoted by ${{\bf{F}}_{(3)}} = {\bf{U}}{\bf \Lambda }{{\bf{V}}^H}$, where ${\bf{U}} \in {\mathbb{C}^{{SKN} \times L}}$, ${\bf{\Lambda}} \in {\mathbb{C}^{L \times L}}$, and ${\bf{V}} \in {\mathbb{C}^{{Q} \times L}}$. According to \textbf{Lemma~\ref{111}}, there must exist a nonsingular matrix $\bf M$ of size $L\times L$ such that
\begin{equation}
{\bf{UM}} = {\bf{K}} \odot {\bf{X}} \odot {\bf B}\label{19}
\end{equation}
or equivalently,
\begin{equation}
{{\bf{U}}_1}{\bf{M}} = {\bf{\overline K}} \odot {\bf{X}}\odot {\bf B}, \qquad {{\bf{U}}_2}{\bf{M}} = {\bf{\underline K}} \odot {\bf{X}}\odot {\bf B}\label{20}
\end{equation}
where submatrices ${{\bf{U}}_1}= \left[ {{{\bf{I}}_{KN(S - 1)}},{{\bf{0}}_{KN(S - 1) \times KN}}} \right]{\bf{U}}$ and ${{\bf{U}}_2}=\left[ {{{\bf{0}}_{KN(S - 1) \times KN}},{{\bf{I}}_{KN(S - 1)}}} \right]{\bf{U}}$ are truncated from rows of $\bf U$. Since $\bf K$ is a Vandermonde matrix, ${\bf{\underline K}} = {\bf{\overline K}}{{\bf \Omega}}$, where ${{\bf \Omega}} = diag({\bm \omega})$. Substitute it into \eqref{20} to obtain
\begin{equation}
{{\bf{U}}_2}{\bf{M}} = {{\bf{U}}_1}{\bf{M}}{{\bf \Omega}}.\label{21}
\end{equation}

Note that $\bf M$ and ${{\bf \Omega}}$ are both full rank, ${{\bf{U}}_2} = {{\bf{U}}_1}\left( {{\bf{M}}{{\bf{\Omega }}}{{\bf{M}}^{ - 1}}} \right)$. After the eigenvalue decomposition (EVD) of the matrix ${\bf{U}}_1^\dag {{\bf{U}}_2}$, ${{\bm{\omega}}}$ can be estimated as the vector of eigenvalues and $\bf M$ is the matrix of the corresponding eigenvectors. Then, the target DOA can be computed by
\begin{equation}
\begin{aligned}
& {\hat \omega}(l) = {e^{ - j\pi {\Delta _m}\sin {\bar \theta _l}}}\\
& {\Delta _m}\sin {\bar \theta _l} - {\Delta _m}\sin {{\bar \theta}^{'}_l} =  \pm 2k \\
\end{aligned}\label{grating}
\end{equation}
where $k \in \left( { - \frac{{\Delta _m }}{2},\frac{{\Delta _m }}{2}} \right)$ is an integer, ${\bar \theta _l}$ is the true direction, and ${{\bar \theta}^{'}_l}$ denotes the potential grating lobes when $\Delta _m \geq 2$.

The estimation of ${\bm\omega}_x$ and ${\bm \omega}_y$ for planar array is straightforward, which can also be found in the proof of \textbf{Lemma~\ref{111}}. Consequently, $\hat u_l$ and $\hat v_l$ can be determined by
\begin{equation}
\left\{ \begin{array}{l}
{{\hat \omega }_y}(l) = {e^{ - j\pi {\Delta _{{m_y}}}{{\bar u}_l}}}\\
{\Delta _{{m_y}}}{{\bar u}_l} - {\Delta _{{m_y}}}{{\bar u'}_l} =  \pm 2{k_y}
\end{array} \right.,\quad \left\{ \begin{array}{l}
{{\hat \omega }_x}(l) = {e^{ - j\pi {\Delta _{{m_x}}}{{\bar v}_l}}}\\
{\Delta _{{m_x}}}{{\bar v}_l} - {\Delta _{{m_x}}}{{\bar v'}_l} =  \pm 2{k_x}
\end{array} \right.\label{grating2}
\end{equation}
where $k_y \in \left( { - \frac{{\Delta _{m_y} }}{2},\frac{{\Delta _{m_y} }}{2}} \right)$ and $k_x \in \left( { - \frac{{\Delta _{m_x} }}{2},\frac{{\Delta _{m_x} }}{2}} \right)$ are integers, respectively, $\bar u_l$ and $\bar v_l$ indicate the DOA information of $l$-th target, while ${{\bar u}^{'}_l}$ and ${{\bar v}^{'}_l}$ correspond to the potential grating lobes. Since $u_l \triangleq \sin\varphi_l\sin\theta_l$ and $v_l \triangleq \sin\varphi_l\cos\theta_l$, the pair of $(\hat \theta_l, \hat\varphi_l)$ can be denoted by
\begin{equation}
{\hat \theta _l} = \arctan \left(\frac{{{\bar u_l}}}{{{\bar v_l}}} \right), \qquad {\bar \varphi _l} = \arcsin \left(\sqrt {\bar u_l^2 + \bar v_l^2} \right). \label{DOA}
\end{equation}

The process in \eqref{19}-\eqref{21} can be regarded as the generalized ESPRIT method\cite{39}. Compared to other tensor-decomposition based methods like PARAFAC, the Vandermonde structure of the factor matrix is exploited and the computational complexity is reduced significantly.  No iterations are required and the convergence is guaranteed.

However, the precondition $r({\bf C}) = L$ must be satisfied. In some applications regarding target detection, it may happen that two targets with similar Doppler shifts exist. Under this circumstance, two column vectors in $\bf C$ are considered to be linearly dependent. The rank deficiency problem limits the application of this computationally efficient DOA estimation method. Besides, the spatial ambiguity problem further restricts the placement of transmit elements. The distance of phase centers between two adjacent subarrays should be no more than half the working wavelength. The array aperture is limited. To tackle the problem of rank deficiency and obtain a higher spatial resolution, \eqref{general} is reshaped by squeezing $\bf B$ and $\bf C$ into one dimension. The third factor matrix ${\bf B} \odot {\bf C}$, as the KR product of a Vandermonde matrix and an arbitrary matrix, generically has rank $\min(QN,L)$.\footnote{Although there
exists no deterministic formula for the rank of the KR product of a Vandermonde matrix and an arbitrary matrix, it is generically full rank. See Appendix~\ref{222}.} Two targets with identical Doppler shift can be resolved, while the grating lobes can be eliminated by comparing the estimation result originated from ${\bf G}$ to the distinct target angular information obtained by ${\bf X}$ \cite{52,53}.

\subsection{Proposed Computationally Efficient DOA Estimation Method for TB MIMO Radar with Uniformly Spaced Transmit Subarrays}\label{key}
Consider the noise-free version of \eqref{general}, a 3-order tensor ${\cal T }\triangleq [[ {{\bf{G}},{\bf{X}},({\bf{B}}\odot{\bf{C}})}]]$ can be reshaped. The mode-3 unfolding of $\cal T$ is given by
\begin{equation}
{{\bf{T}}_{(3)}} = \left( {{\bf{G}} \odot {\bf{X}}} \right){\left( {{\bf{B}} \odot {\bf{C}}} \right)^T}\label{T3}
\end{equation}
where $\bf G$, $\bf X$, and ${\bf{B}} \odot {\bf{C}}$ are the three factor matrices, respectively. The receive steering matrix and Doppler steering matrix are squeezed into one dimension. Note that the generators of $\bf G$ are distinct, the directions of all targets are unique with or without the existence of grating lobes. Hence, the third factor matrix ${\bf{B}} \odot {\bf{C}}$ is column full rank \cite{29}. \textbf{Lemma~\ref{111}} holds for tensor $\cal T$. In the following, we develop methods for DOA estimation in TB MIMO radar with uniformly spaced subarrays via the decomposition of $\cal T$ for linear and planar array sequentially.

\subsubsection{ULA}
Let $\bf G = \bf K$. According to \textbf{Lemma~\ref{111}}, the decomposition of $\cal T$ is unique. To obtain the factor matrices with target DOA information, denote the SVD of ${{\bf{T}}_{(3)}}$ as ${{\bf{T}}_{(3)}} = {\bf{U}}{\bf \Lambda }{{\bf{V}}^H}$, where ${\bf{U}} \in {\mathbb{C}^{{SK} \times L}}$, ${\bf{\Lambda}} \in {\mathbb{C}^{L \times L}}$, and ${\bf{V}} \in {\mathbb{C}^{{NQ} \times L}}$. A nonsingular matrix $\bf E$ of size $L\times L$ satisfies
\begin{equation}
{\bf{UE}} = {\bf{K}} \odot {\bf{X}}.\label{UE}
\end{equation}

Owing to the operator of the KR product, we can write
\begin{equation}
{{\bf{U}}_1}{\bf{E}} = {\bf{\overline K}} \odot {\bf{X}}, \qquad {{\bf{U}}_2}{\bf{E}} = {\bf{\underline K}} \odot {\bf{X}}\label{rr}
\end{equation}
where ${{\bf{U}}_1}= \left[ {{{\bf{I}}_{K(S - 1)}},{{\bf{0}}_{K(S - 1) \times K}}} \right]{\bf{U}}$ and ${{\bf{U}}_2}=\left[ {{{\bf{0}}_{K(S - 1) \times K}},{{\bf{I}}_{K(S - 1)}}} \right]{\bf{U}}$ are truncated from rows of $\bf U$, respectively. Substitute ${\bf{\underline K}} = {\bf{\overline K}}{{\bf \Omega}}$ into \eqref{rr} to obtain
\begin{equation}
{{\bf{U}}_2}{\bf{E}} = {{\bf{U}}_1}{\bf{E}}{{\bf \Omega}}.\label{UK}
\end{equation}

Since $\bf E$ and ${{\bf \Omega}}$ are both full rank, ${{\bf{U}}_2} = {{\bf{U}}_1}\left( {{\bf{E}}{{\bf{\Omega }}}{{\bf{E}}^{ - 1}}} \right)$. The connection between ${\bf{U}}_1^\dag {{\bf{U}}_2}$ and ${{\bf{E}}{{\bf{\Omega }}}{{\bf{E}}^{ - 1}}}$ is revealed. From the EVD of ${\bf{U}}_1^\dag {{\bf{U}}_2}$, the generators ${{\bm{\omega}}}$ can be estimated with $\bf E$ being the matrix of the corresponding eigenvectors. Then, $\left\{ {{\hat \theta _{l}}} \right\}_{l = 1}^L$ can be computed by \eqref{grating}.

Note that $\left( {\frac{{{\bm{\kappa }}_l^H}}{{{\bm{\kappa }}_l^H{{\bm{\kappa }}_l}}} \otimes {{\bf{I}}_K}} \right)\left( {{{\bm{\kappa }}_l} \otimes {{\bm{\chi }}_l}} \right) = {{\bm{\chi }}_l}$ and ${{\bm{\kappa }}_l^H}{{\bm{\kappa }}_l} = S$, the compact form of ${{\bm{\chi }}_l}$ is given as
\begin{equation}
{{\bm{\chi }}_l} = 1/S \left( {{{\bm{\kappa }}_l^H} \otimes {{\bf{I}}_K}} \right){\bf{U}}{{\bf{e}}_l}.\label{K1}
\end{equation}

Equation~\eqref{K1} provides an estimation of each column vector of $\bf X$. Given ${\bf W}_0$ as prior information, a polynomial rooting method \cite{52} can be applied to estimate the unambiguous $\left\{ {{\theta _{l}}} \right\}_{l = 1}^L$ in ${\bf A}_0$ independently. Instead of exploiting the signal subspace shift-invariance of the transmit subarray steering matrix, the method in \cite{52} focuses on the Vandermonde structure of ${\bf{A}}_0$ within a single subarray and reveals the relationship between the TB MIMO radar transmit beampattern and the generalized sidelobe canceller (GSC). Consequently, the estimation results originated from ${\bf K}$ and ${\bf X}$ both provide the target angular information. The grating lobes can be eliminated by comparing the results to each other. Note that \eqref{K1} is conducted column by column, the angles are paired automatically before comparison.

An outline of the proposed method for the DOA estimation in TB MIMO radar with linear array is given as \textbf{Algorithm~\ref{alg1}}.
\begin{algorithm}[tb]
\caption{DOA Estimation for 1-D TB MIMO Radar with Uniformly Spaced Transmit Subarrays} \label{alg1}
\begin{algorithmic}[1]
\REQUIRE ~~\\
Signal tensor ${\cal{Z}} \in {\mathbb{C}^{S \times K \times N \times Q}}$ from \eqref{tensor1}
\ENSURE ~~\\
Targets DOA information $\left\{ {{\theta _{l}}} \right\}_{l = 1}^L$
\STATE Reshape ${\cal{Z}}$ into a 3-order tensor ${\cal{T}} \in {\mathbb{C}^{{S} \times K \times {NQ}}}$, where the mode-3 unfolding of ${\cal{T}}$ is given by \eqref{T3};
\STATE Compute the SVD of the matrix ${{\bf{T}}_{(3)}} = {\bf{U}}{\bf \Lambda }{{\bf{V}}^H}$;
\STATE Formulate two submatrices ${{\bf{U}}_1},{{\bf{U}}_2}$ satisfying \eqref{rr};
\STATE Calculate the EVD of the matrix ${\bf{U}}_1^\dag {{\bf{U}}_2}$;
\STATE Estimate ${\hat \theta}_l$ via \eqref{grating}, which contains grating lobes;
\STATE Construct ${{\bm{\chi }}_l}$ via \eqref{K1};
\STATE Define ${\bf{\tilde W}_0}  \triangleq {\bf{W}}_0- {\bf W}'_0$,  ${{\bf{W}}'_0} \triangleq {\left[ {{{\bm{\chi}}_l},{{\bf{0}}_{K \times \left( {{\rm{{M_0} - 1}}} \right)}}} \right]^T}$;
\STATE Build a polynomial via $F({z_l}) \triangleq {{\bf{p}}^H}({z_l}){\bf{\tilde W}}_0{{{\bf{\tilde W}}}_0^H}{\bf{p}}({z_l})$, where ${\bf{p}}(z_l) \triangleq {\left[ {1,z_l,\cdots,{{z_l}^{{M_0} - 1}}} \right]^T}$ and $z_l \triangleq e^{-j\pi\sin\theta_l}$;
\STATE Compute the roots of the polynomial $F({z_l})$ and select the one closest to the unit circle as $\hat z_l$;
\STATE Estimate ${\theta}_l $ via ${\hat \theta}_l = \arcsin \left(\frac{{j\ln ({\hat z_l})}}{\pi }\right)$;
\STATE Compare the results in step~5 and step~10;
\RETURN $\left\{ {{\theta _{l}}} \right\}_{l = 1}^L$.
\end{algorithmic}
\end{algorithm}

\subsubsection{URA}
First, substitute $ {\bf G } = {\bf H} \odot {\bf \Delta}$ into \eqref{T3}. Similar to \eqref{UE}, the SVD of ${{\bf{T}}_{(3)}}$ is ${{\bf{T}}_{(3)}} = {\bf{U}}{\bf \Lambda }{{\bf{V}}^H}$, and there is a nonsingular matrix ${\bf E} \in {\mathbb{C}^{ L \times L}}$ such that
\begin{equation}
{\bf{UE}} = {\bf H} \odot {\bf \Delta} \odot {\bf{X}}.
\end{equation}

Considering the KR product, the Vandermonde structure of both ${\bf H}$ and ${\bf \Delta}$ is exploited via
\begin{equation}
\begin{aligned}
& {{\bf{U}}_{\rm{2}}}{\bf{E}} = {\bf{\underline H }} \odot {\bf{\Delta}} \odot {\bf{X}} = \left( {{\bf{\overline H }} \odot {\bf{\Delta}} \odot {\bf{X}}} \right){{\bf{\Omega }}_y} = {{\bf{U}}_1}{\bf{E}}{{\bf{\Omega }}_y}\\
& {{\bf{U}}_{\rm{4}}}{\bf{E}} = {\bf{H }} \odot {\bf{\underline \Delta}} \odot {\bf{X}} = \left( {{\bf{H }} \odot {\bf{\overline \Delta}} \odot {\bf{X}}} \right){{\bf{\Omega }}_x} = {{\bf{U}}_3}{\bf{E}}{{\bf{\Omega }}_x}\\
\end{aligned}
\end{equation}
where ${{\bf \Omega}_y} = diag({{\bm\omega}_y})$, ${{\bm \Omega}_x} = diag({{\bm\omega}_x})$, ${{\bf{U}}_{\rm{1}}}$, ${{\bf{U}}_{\rm{2}}}$, ${{\bf{U}}_{\rm{3}}}$ and ${{\bf{U}}_{\rm{4}}}$ are the submatrices truncated from rows of ${\bf{U}}$, i.e.,
\begin{equation}
\begin{aligned}
&{{\bf{U}}_1}{\rm{ = }}\left[ {{{\bf{I}}_{{I}K({J} - 1)}},{{\bf{0}}_{{I}K({J} - 1) \times {I}K}}} \right]{\bf{U}}\\
&{{\bf{U}}_2}{\rm{ = }}\left[ {{{\bf{0}}_{{I}K({J} - 1) \times {I}K}},{{\bf{I}}_{{I}K({J} - 1)}}} \right]{\bf{U}}\\
&{{\bf{U}}_3} = \left( {{{\bf{I}}_{{J}}} \otimes \left[ {{{\bf{I}}_{K({I} - 1)}},{{\bf{0}}_{K({I} - 1) \times K}}} \right]} \right){\bf{U}}\\
&{{\bf{U}}_4} = \left( {{{\bf{I}}_{{J}}} \otimes \left[ {{{\bf{0}}_{K({I} - 1) \times K}},{{\bf{I}}_{K({I} - 1)}}} \right]} \right){\bf{U}}.\\
\end{aligned}\label{select}
\end{equation}

Like \eqref{UK}, the vectors ${{\bm{\omega}}_y}$ and ${{\bm{\omega}}_x}$ can be estimated as the collections of eigenvalues of ${\bf{U}}_1^\dag {{\bf{U}}_2}$ and ${\bf{U}}_3^\dag {{\bf{U}}_4}$, respectively, and $\bf E$ is the matrix of the corresponding eigenvectors. Then, the possible pairs of $(\theta_l, \varphi_l)$ can be computed by \eqref{grating2}-\eqref{DOA}. To eliminate the grating lobes, the relationship between the TB MIMO radar transmit beampattern and the GSC is applied again to estimate the target DOA in 2-D case \cite{53}. Specifically, $\left( {\frac{{{\bm{\kappa }}_l^H}}{{{\bm{\kappa }}_l^H{{\bm{\kappa }}_l}}} \otimes {{\bf{I}}_K}} \right)\left( {{{\bm{\kappa }}_l} \otimes {{\bm{\chi }}_l}} \right) = {{\bm{\chi }}_l}$ and ${{\bm{\kappa }}_l^H}{{\bm{\kappa }}_l} = S$ still hold by replacing ${{\bm{\kappa }}_l}$ with ${{\bm{h}}_l} \odot {{\bm{\delta}}_l}$. Hence, each column of $\bf X$ can be restored by
\begin{equation}
{{\bm{\chi }}_l} = 1/S\left[ {({{\bm{h}}_l} \odot {{\bm{\delta}}_l})^H \otimes {{\bf{I}}_K}} \right]{\bf{U}}{{\bf{e}}_l}.\label{K}
\end{equation}

Note that the TB matrix $ {\bf{W}}_0$ is given as a prior information, ${{\bm{\chi }}_l} = {\bf{W}}_0^H{\bm{\alpha }}_0(\theta_l ,\varphi_l )$ can be rewritten as $K$ different linear equations
\begin{equation}
{{{\chi }}_l}(k) = {\bf{w}}_k^H{\bm{\alpha }}_0(\theta_l ,\varphi_l ), \quad k = 1,2,\cdots, K \label{element}
\end{equation}
or equivalently, ${\bf{p}}_k^H{\bm{\alpha }}_0(\theta_l ,\varphi_l ) = 0$, where ${\bf{p}}_k \triangleq {\bf{w}}_k-[{{{\chi }}_l}(k) , {\bf 0}_{1 \times (M-1)}]$. It can be seen that the linear equations in \eqref{element} hold if and only if ${{{\left|\left| {{\bf{P}}^H{{\bf{a }}}({\hat \theta _l},{\hat\phi _l})}\right|\right|^2}}} = 0$, where ${\bf P} = {\bf W}-{\bf W}_0$, ${\bf W}_0 \triangleq [{{{\bm \chi}}_{l}}, {\bf 0}_{K \times (M-1)}]^T$. Therefore, the estimation of a pair $(\theta_l ,\varphi_l )$ can be found by solving the following convex optimization problem \cite{53}
\begin{equation}
\begin{aligned}
&\min \limits_{(\hat \theta_l,\hat \phi_l)} {{{\left|\left| {{\bf{P}}^H({\bf{u}}(\hat \theta_l ,\hat \varphi_l ) \otimes {\bf{v}}(\hat \theta_l ,\hat \varphi_l ))}\right|\right|^2}}}.\label{convex}
\end{aligned}
\end{equation}
whose structure is similar with the TB MIMO radar transmit beampattern. After obtaining the pair $(\hat u_l, \hat v_l)$, the distinct target DOA can be computed by \eqref{DOA}. The computation is conducted via \eqref{K} column by column, hence the independent estimates of target DOA from $\bf G$ and $\bf X$ are paired automatically. By comparing them,  the grating lobes can be mitigated.

The primary procedures for the DOA estimation in TB MIMO radar with planar array is summarized as \textbf{Algorithm~\ref{alg2}}.
\begin{algorithm}[tb]
\caption{DOA Estimation for 2-D TB MIMO radar with Uniformly Spaced Transmit Subarrays} \label{alg2}
\begin{algorithmic}[1]
\REQUIRE ~~\\
Signal Tensor ${\cal{Z}} \in {\mathbb{C}^{{J}\times {I} \times K \times N \times Q}}$ from \eqref{tensor2}
\ENSURE ~~\\
Targets DOA information $\left\{ {{\theta _{l}}} \right\}_{l = 1}^L$ and $\left\{ {{\varphi _{l}}} \right\}_{l = 1}^L$
\STATE Reshape ${\cal{Z}}$ into a 3-order tensor ${\cal{T}} \in {\mathbb{C}^{{S} \times K \times {NQ}}}$, where the mode-3 unfolding of ${\cal{T}}$ is given by \eqref{T3};
\STATE Compute the SVD of the matrix ${{\bf{T}}_{(3)}} = {\bf{U}}{\bf \Lambda }{{\bf{V}}^H}$;
\STATE Formulate four submatrices ${{\bf{U}}_1},{{\bf{U}}_2},{{\bf{U}}_3},{{\bf{U}}_4}$ via \eqref{select};
\STATE Calculate the EVD of the matrices ${\bf{U}}_1^\dag {{\bf{U}}_2}$ and ${\bf{U}}_3^\dag {{\bf{U}}_4}$;
\STATE Estimate the pair $({\hat u}_l,\hat v_l)$ via \eqref{grating2} and compute the pair $(\hat \theta_l,\hat \phi_l)$ via \eqref{DOA};
\STATE Construct ${{\bm{\chi }}_l}$ via \eqref{K}, where $\bf E$ is the matrix of the corresponding eigenvectors in step~4;
\STATE Build the matrix $\bf P$ and solve the minimization problem \eqref{convex} to obtain the pair $(\hat u_l,\hat v_l)$;
\STATE Compute the unambiguous pair $(\hat \theta_l, \hat \varphi_l)$ via \eqref{DOA};
\STATE Compare the results in step~5 and step~8;
\RETURN $\left\{ {{\theta _{l}}} \right\}_{l = 1}^L$ and $\left\{ {{\varphi _{l}}} \right\}_{l = 1}^L$.
\end{algorithmic}
\end{algorithm}

\subsection{Parameter Identifiability}
As mentioned in \textbf{Fact~\ref{krank}}, the generical uniqueness condition of tensor decomposition for a high-order tensor is given as $\sum\limits_{n = 1}^N {{k_{{{\bf{A}}^{(n)}}}}}  \ge 2L + (N - 1)$, where $N$ is the tensor order and $L$ is the tensor rank. The upper bound of the tensor rank, i.e., the maximum number of targets that can be resolved, rises with the increase of tensor order at the level of Kruskal-rank of a matrix. However, if the factor matrix has special structure, the uniqueness condition is changed. An example of tensor decomposition with Vandermonde factor matrix is described in \textbf{Lemma~\ref{111}}. It can be observed that the maximum number of targets that can be resolved by \eqref{general} is determined by the preconditions. The first precondition is that the first factor matrix of the tensor, as a Vandermonde matrix or the KR product of two Vandermonde matrices, must have distinct vector of generators. In this paper, we assume this condition holds since it means that each of the targets posses a unique direction, which is reasonable regarding DOA estimation problem.

The second precondition requires that the third factor matrix has rank $L$. Note that the Doppler steering vectors of any two targets with the same Doppler shift are linearly dependent with a scale difference determined by the target RCS. Two scenarios are discussed next.

When the target Doppler shifts are unique, \textbf{Lemma~\ref{111}} can be applied directly and the tensor $\cal F$ can be used. To ensure the uniqueness decomposition, it is required that \cite{29}
\begin{equation}
\min ((S-1)KN, Q) \geq L.
\end{equation}

In MIMO radar, the number of pulses during a single CPI is usually large. Thus, the maximum number of targets that can be resolved is generically $(S-1)KN$, which is better than that in \textbf{Fact~\ref{krank}} \cite{29}. However, the size of the transmit array is confined since the distance between phase centers of two adjacent subarrays must be no more than half the working wavelength to avoid the spatial ambiguity. This restriction degrades the spatial resolution and also raises the difficulty of the physical implementation of the array.

When there are at least two targets that have identical Doppler shift, tensor $\cal T$ is used to ensure the second precondition. The receive steering matrix is squeezed together with the Doppler steering matrix to distinguish targets with identical velocity. Although the rank of a specific tensor remains the same when it is reshaped, it was proved that different reshape manners are not equivalent from the performance point of view \cite{47}. In our case, it means that the identifiability is changed and, therefore, the uniqueness condition of decomposition for $\cal T$ requires
\begin{equation}
\min \left((S-1)K, NQ\right) \geq L.
\end{equation}

The usage of $\cal T$ is more appropriate for the general case, since the rank deficiency problem caused by identical target Doppler shift is solved. Additionally, by reshaping tensor $\cal Z$ into tensor $\cal T$, the angular information can be estimated independently from $\bf G$ and $\bf X$. The unambiguous estimation result from $\bf X$ provides a second estimation of the target DOA and can be used to eliminate the grating lobes. Thus, it can be concluded that the use of the tensor model $\cal T$ has at least the above two advantages. The maximum number of targets that can be resolved is reduced to $(S-1) \cdot K$. To improve the parameter identifiability, the increase of number of transmit subarrays or transmit waveforms is worth considering. 

\section{Arbitrary but Identical Subarrays with Multiple Scales of Shift-invariances }\label{4}
In previous section, we have assumed that the transmit subarrays are uniformly spaced to obtain a Vandermonde structure in the factor matrix of designed tensor. However, such constraint on subarray structure can be relaxed. The placement of all subarrays needs not be uniform, while the configuration within a single subarray can be arbitrary. The tensor model in \eqref{general} is applicable for TB MIMO radar with any arbitrary but identical subarrays, since the extended factor matrix that represents the phase rotations between transmit subarrays is merely determined by the coordinates of the transmit subarray phase centers. The difference is that the array configuration varies the structure of the factor matrix, which may cause extra steps to recover the target DOAs. A typical example has been given earlier where the unambiguous spatial information in $\bf X$ is exploited to eliminate the cyclic ambiguity in $\bf G$.

In the following, we discuss two general cases that the transmit array with multiple scales of shift-invariances is placed on a lattice and explain the use of the proposed computationally efficient DOA estimation method in both scenarios.

\subsection{Generalized Vandermonde Matrix}
Note that the Vandermonde structure of the steering matrix ${\bf G}$ is linked to the phase rotations between the transmit subarrays, and it is exploited in a look up table for finding target DOAs. The Vandermonde structure is only a special case leading to phase rotation. Indeed, take, for example a linear array with its elements placed on a lattice, where all lattice cells are enumerated sequentially. The indices of the elements form a counted set of increasing positive integers. It can be shown that $m_s$ must be a subset of this set, since the first element of each subarray corresponds to a unique lattice cell. Hence, $m_s$ may increase uniformly or non-uniformly.

From \eqref{generate}, it can be observed that $m_s$ determines ${\bf K}$. When it rises uniformly, ${\bf K}$ should be a Vandermonde matrix.\footnote{This has been derived in Section~\ref{sig}, and we can find that the shift-invariance between different subarrays is related to the step size of $m_s$, or more specifically, the coordinates of the transmit subarray phase centers.} Determined by the step size of $m_s$, i.e., $\Delta_m$, the configuration of adjacent subarrays can be partly overlapped ($\Delta_m < M_0$) or non-overlapped ($\Delta_m \ge M_0$). The proposed DOA estimation method in last section can be used directly.

In the case when $m_s$ rises non-uniformly, let us consider as an example the tensor model \eqref{tensor1} with $S = 7$ subarrays and $m_s = \{1,2,3,5,6,7,9\}$. Each subarray contains three elements, therefore, the original transmit array is a ULA with $M = 11$ elements. Then, $\bf K$ is a {\it generalized Vandermonde matrix} \cite{32}, which can be written as ${\bf K} \triangleq [{\bf z}_1,\cdots, {\bf z}_L]$, where ${\bf z}_l \triangleq [1,z_l,z_l^2,z_l^4,z_l^5,z_l^6,z_l^8]^T$ and ${z_l} \triangleq  {e^{ - j\pi \sin {\theta _l}}}$.

The idea of multiple invariance ESPRIT\cite{50} is introduced to conduct the DOA estimation with non-uniformly spaced transmit subarrays. Consequently, $\bf  K$ can be interpreted as the combination of a set of submatrices ${\bf K}^{(sub)}$ denoting different sub-ULAs associated with various shift-invariances, i.e.,
\begin{equation}
\begin{aligned}
&{\bf K}^{(sub)} \triangleq \left[\left({\bf K}^{(1,1)}\right)^T, \left({\bf K}^{(2,1)}\right)^T,\left({\bf K}^{(1,2)}\right)^T\right]^T\\
&{\bf K}^{(1,1)}\triangleq \left[{\bf z}^{(1,1)}_1,\cdots, {\bf z}^{(1,1)}_L\right]\\
&{\bf K}^{(2,1)} \triangleq \left[{\bf z}^{(2,1)}_1,\cdots, {\bf z}^{(2,1)}_L\right]\\
&{\bf K}^{(1,2)} \triangleq \left[{\bf z}^{(1,2)}_1,\cdots, {\bf z}^{(1,2)}_L\right]
\end{aligned}\label{case}
\end{equation}
where ${\bf z}^{(1,1)}_l$ is selected from ${\bf z}_l$ with $m_s = \{1,2,3\}$, ${\bf z}^{(2,1)}_l$ is selected from ${\bf z}_l$ with $m_s = \{5,6,7\}$, and ${\bf z}^{(1,2)}_l$ is selected from ${\bf z}_l$ with $m_s = \{1,3,5,7,9\}$. In other words, ${\bf K}^{(1,1)}$ is a submatrix of $\bf K$ that consists of first three rows with shift-invariance $\Delta_m = 1$. The other two submatrices are analogous. Note that \eqref{case} is not the only subarray construction method, but it contains all transmit subarrays with a minimal distinct shift-invariance set $\Delta = \{\Delta_m | \Delta_m = 1,2\}$.

Substituting \eqref{case} to \eqref{T3}, we can write
\begin{equation}
{{\bf{T}}^{(sub)}_{(3)}} = \left( {{\bf{K}}^{(sub)} \odot {\bf{X}}} \right){\left( {{\bf{B}} \odot {\bf{C}}} \right)^T}.\label{Tsub}
\end{equation}
Its SVD is given as ${{\bf{T}}^{(sub)}_{(3)}} = {\bf U}^{(sub)}{\bf \Lambda}^{(sub)}\left({\bf V}^{(sub)}\right)^H$. It can be observed that \textbf{Lemma~\ref{111}} holds for \eqref{Tsub}. By constructing ${\bf K}^{(sub)}$, a new transmit subarray steering matrix that consists of several Vandermonde submatrices can be introduced. To exploit the Vandermonde structure, an extra row selection must be applied. Taking ${\bf{K}}^{(1,1)}$, for example, we can generalize \eqref{UE} to obtain
\begin{equation}
{{\bf{K}}^{(1,1)} \odot {\bf{X}}} = {\bf U}^{(1,1)}{\bf E}\label{40}
\end{equation}
where ${\bf U}^{(1,1)}$ is truncated from ${\bf U}^{(sub)}$ in the same way as ${\bf K}^{(1,1)}$ from ${\bf K}^{(sub)}$. Thus, each column of ${\bf K}^{(1,1)}$ can be estimated. The estimates of ${\bf K}^{(1,2)}$ and ${\bf K}^{(2,1)}$ can be obtained similarly. It is worth noting that if $\Delta_m >1$, the problem of grating lobes may still occur when recovering $\theta_l$ from  ${\bf U}^{\left({N_{\Delta_m}},\Delta_m\right)}$, where $N_{\Delta_m}$ represents the number of subarrays whose shift-invariance is determined by $\Delta_m$. Usually, the unambiguous spatial information in $\bf X$ can be exploited to eliminate the potential grating lobes. However, it requires each subarray to be dense ULA, which restricts the aperture of the transmit subarray, and therefore, the spatial resolution.

Note that the generators of the Vandermonde submatrices in ${\bf K}^{(sub)}$ provide the target DOA information at different exponential levels, i.e., $z_l^{\Delta_m}$. Based on this, a polynomial function is designed to estimate the target DOA without using the second factor matrix ${\bf X}$.

For every possible shift-invariance $\Delta_m$, denote
\begin{equation}
\begin{aligned}
&{\bf a}_l^{(\Delta_m)} \triangleq \left[{ \overline {\bm \kappa}}_l^{(1, \Delta_m)T},\cdots, {\overline {\bm \kappa}}_l^{({N_{\Delta_m}}, \Delta_m)T}\right]^T\\
&{\bf b}_l^{(\Delta_m)} \triangleq \left[{\underline {\bm \kappa}}_l^{(1, \Delta_m)T},\cdots, {\underline {\bm \kappa}}_l^{({N_{\Delta_m}}, \Delta_m)T}\right]^T.\\
\end{aligned}\label{41}
\end{equation}

To illustrate \eqref{41}, consider the array structure in \eqref{case}. When $\Delta_m = 1$, there are two different submatrices/sub-ULAs, i.e., ${N_{1}} = 2$, ${\bf a}_l^{(1)} = \left[{ \overline {\bm \kappa}}_l^{(1, 1)T},{ \overline {\bm \kappa}}_l^{(2, 1)T}\right]^T = \left[1, z_l, z_l^4,z_l^5\right]^T$, and ${\bf b}_l^{(1)} = \left[{ \underline {\bm \kappa}}_l^{(1, 1)T},{ \underline {\bm \kappa}}_l^{(2, 1)T}\right] ^T= \left[z_l, z^2_l, z_l^5,z_l^6\right]^T$. When $\Delta_m = 2$, only one submatrix/sub-ULA exists, i.e., ${N_{2}} = 1$, ${\bf a}_l^{(2)} = { \overline {\bm \kappa}}_l^{(1,2)} = \left[1, z_l^2, z_l^4,z_l^6\right]^T$ and ${\bf b}_l^{(2)} = { \underline{\bm \kappa}}_l^{(1,2)} = \left[z_l^2, z_l^4,z_l^6,z_l^8\right]^T$. Also, the following constraint should be satisfied
\begin{equation}
{\bf a}_l^{(\Delta_m)}{{z}_l^{\Delta_m}} = {\bf b}_l^{(\Delta_m)}.\label{42}
\end{equation}

It is proved in \cite{32} that \eqref{42} can be achieved by rooting the polynomial function
\begin{equation}
f({z_l}) \triangleq \sum\limits_{{\Delta _m} \in \Delta } {\left| {\left| {{\bf{a}}_l^{({\Delta _m})}z_l^{{\Delta _m}} - {\bf{b}}_l^{({\Delta _m})}} \right|} \right|_F^2} \label{43}
\end{equation}
as long as two coprime numbers  can be found in the shift-invariance set $\Delta$. By definition of $z_l$, the root nearest to the unit circle should be chosen as $\hat z_l$, which finally estimates the target DOA as ${\hat \theta}_l = \arcsin \left(\frac{{j\ln ({\hat z_l})}}{\pi }\right)$. The construction of ${\bf K}^{(sub)}$ enables the use of $\cal T$ in a more general scenario. The transmit subarrays can be organized in a non-uniform way. If the shift-invariance set $\Delta$ contains a pair of coprime integers, the problem of spatial ambiguity can be solved with no limitation on the transmit subarray structure. Hence, the structures of the transmit subarrays can be arbitrary but identical.

An outline of the proposed DOA estimation method for TB MIMO radar with non-uniformly spaced arbitrary but identical subarrays is summarized in \textbf{Algorithm~\ref{alg3}}.
\begin{remark}:
Multiple scales of shift-invariances can also be found in a Vandermonde matrix ${\bf K}$\cite{50}. A simple way to build ${\bf K}^{(sub)}$ is to concatenate the submatrices of ${\bf K}$, which, respectively, consist of the odd rows, even rows and all rows. Hence, \textbf{Algorithm~\ref{alg3}} is also applicable for TB MIMO radar with uniformly spaced transmit subarrays. Since the manifold of the subarray does need not to be dense ULA, it is possible to place the transmit array on a larger lattice to obtain a higher spatial resolution. If some elements in a transmit subarray are broken, a useful solution is to disable the elements in other subarrays accordingly to keep the manifolds identical. Moreover, we can select part of the elements in all subarrays to fulfill other purposes like communication in joint radar-communication system for example \cite{56}. These remarks can be extended to the case of planar array.
\end{remark}

\begin{algorithm}[tb]
\caption{DOA Estimation for TB MIMO radar with Non-Uniformly Spaced Arbitrary but Identical Subarrays} \label{alg3}
\begin{algorithmic}[1]
\REQUIRE ~~\\
Signal Tensor ${\cal{Z}} \in {\mathbb{C}^{S \times K \times N \times Q}}$ from \eqref{general}
\ENSURE ~~\\
Targets DOA information $\left\{ {{\theta _{l}}} \right\}_{l = 1}^L$
\STATE Construct a new matrix ${\bf K}^{(sub)}$ in \eqref{case}, which can be divided into several Vandermonde submatrices and contains all transmit subarrays;
\STATE Update the transmit subarray steering matrix and reshape ${\cal{Z}}$ into a 3-order tensor ${\cal{T}} \in {\mathbb{C}^{{S'} \times K \times {NQ}}}$;
\STATE Compute the SVD of the matrix ${{\bf{T}}^{(sub)}_{(3)}} = {\bf U}^{(sub)}{\bf \Lambda}^{(sub)}{\bf V}^{(sub)H}$;
\STATE Estimate each Vandermonde submatrix in ${\bf K}^{(sub)}$ sequentially via \eqref{UE}-\eqref{UK};
\STATE Build two vectors ${\bf a}_l^{(\Delta_m)}$ and ${\bf b}_l^{(\Delta_m)}$ from \eqref{41} for each column of estimated ${\bf K}^{(sub)}$;
\STATE Compute the roots of the polynomial function in \eqref{43} and select the root nearest to the unit circle as $\hat z_l$;
\STATE Estimate  ${\hat \theta}_l = \arcsin \left(\frac{{j\ln ({\hat z_l})}}{\pi }\right)$;
\RETURN $\left\{ {{\theta _{l}}} \right\}_{l = 1}^L$.
\end{algorithmic}
\end{algorithm}

\subsection{Multiscale Sensor Array}\label{scale}
Another case when multiple scales of shift-invariances can be found is called multiscale sensor array \cite{43,45,46}. Generally, a URA can be regarded as a 2-level multiscale sensor array with different scales of shift-invariances. As shown in Fig.~\ref{sys}, the generation process of such a URA contains two steps. First, consider a single subarray composed of $M_0$ elements as the reference subarray. Let $I$ replica subarrays be placed uniformly across the $x$-axis, which form a larger subarray at a higher level. Then, $J$ copies of this higher level subarray are organized uniformly across the $y$-axis. Combining them together, a URA is constructed. Note that in this specific case, $I$ subarrays at level-1 are non-overlapped, while their $J$  counterparts at level-2 are partly overlapped.

From \eqref{tensor2}, it is clear that the transmit subarray steering matrices for subarrays at level-1 and level-2 are $\bf \Delta$ and $\bf H$, respectively. If the URA itself is repeated and spatially moved to other arbitrary but known locations, a new array that yields a larger spatial aperture is created. In this way, an $R$-level multiscale sensor array can be constituted. For any $S_r$ subarrays at level-$r$, $r = 1,2,\cdots,R$, define ${\bf G}^{(r)} \in {\mathbb{C}^{{S_r} \times L}}$ as the transmit subarray steering matrix. The overall transmit subarray steering matrix is given by
\begin{equation}
{\bf G} = {\bf G}^{(R)} \odot {\bf G}^{(R-1)} \odot \cdots \odot {\bf G}^{(1)} \triangleq \mathop  \odot \limits_{r = 1}^R {{\bf{G}}^{(r)}}.\label{G}
\end{equation}

Substituting \eqref{G} into \eqref{general}, the tensor model for TB MIMO radar with an $R$-level miltiscale sensor array at transmit side is given. Note that this is an $(R+3)$-order tensor. The reshape of it and the use of \textbf{Lemma~\ref{111}} in this case are quite flexible. The DOA estimation can be conducted via \textbf{Algorithm~\ref{alg2}} or \textbf{Algorithm~\ref{alg3}} analogously.

Take a cubic transmit array,\footnote{Repeat the URA in Fig.~\ref{sys} $D$ times across the $z$-axis with coordinates $(0,0,m_d), d = 1,\cdots, D$.} for example. It is a 3-level multiscale sensor array, and we can immediately write the three transmit subarray steering matrices as ${\bf G}^{(1)} = {\bf\Delta}$, ${\bf G}^{(2)} = {\bf H}$ and ${\bf G}^{(3)} = {\bf \Xi}$, respectively, where ${\bf \Xi} \triangleq {\left[{\bm \tau}_1,{\bm\tau}_2,\cdots, {\bm \tau}_D\right]^T}_{ V\times L}$ and ${\bm\tau}_d \triangleq \left[e^{-j\pi(m_d-1)cos\varphi_1},e^{-j\pi(m_d-1)cos\varphi_2},\cdots, e^{-j\pi(m_d-1)cos\varphi_L}\right]^T$.

The parameter identifiability for different reshaped 3-order tensors ${\cal T} \in {\mathbb{C}^{{I_1} \times {I_2} \times {I_3}}}$ varies, which is determined by the uniqueness condition of tensor decomposition, i.e.,
\begin{equation}
\min ((I_1-1)I_2, I_3) \geq L \label{45}
\end{equation}
where $I_1$, $I_2$ and $I_3$ can be regarded as permutations of the set $\{S_1,S_2,\cdots,S_R,K,N,Q\}$, and $I_1I_2I_3 = KNQ\prod\limits_{r = 1}^R {{S_r}}$. From \eqref{45}, it is possible to estimate the target DOAs using only one single pulse. The transmit array with multiple scales of shift-invariances is exploited via the tensor reshape to make up for the lack of number of snapshots. This property can also be used to distinguish coherent sources. An example of two targets with identical Doppler shift has been discussed in Section~\ref{key}. See \cite{51} for more discussions about the partial identifiability of the tensor decomposition, where specific conditions for coherent or collocated sources are investigated.

\section{Simulation Results}\label{5}
In this section, we investigate the DOA estimation performance of the proposed method in terms of the root mean square error (RMSE) and probability of resolution of closely spaced targets for TB MIMO radar. Throughout the simulations, there are $Q = 50$ pulses in a single CPI. We assume that $L = 3$ targets lie within a given spatial steering vector determined by $\left\{ {{\theta _{l}}} \right\}_{l = 1}^L$ in linear array and $\left\{\left( {{\theta _{l}}}, {\varphi_l} \right) \right\}_{l = 1}^L$ in planar array, the normalized Doppler shifts are $f_1 = -0.1, f_2 = 0.2$ and $ f_3 = 0.2$. The number of Monte Carlo trials is $P = 200$. The RCS of every target is drawn from a standard Gaussian distribution, and obeys the Swerling I model. Note that the last two targets share identical Doppler shift, which cause $\bf C$ to drop rank. The noise signals are assumed to be Gaussian, zero-mean and white both temporally and spatially. The $K$ orthogonal waveforms are ${S_k}(t) = \sqrt {\frac{1}{{{T}}}} {e^{j2\pi \frac{k}{{{T}}}t}}, \, k=1, \cdots, K$. For both linear and planar array, the tensor model in \eqref{general} is used and the TB matrix is pre-designed \cite{12,31}.

For linear array, we assume a transmit ULA with $S = 8$ subarrays. Each transmit subarray has $M_0 = 10$ elements spaced at half the wavelength. The placement of transmit subarrays can vary from totally overlapped case to non-overlapped case. The number of transmit elements is computed by $M = {M_0}+{{\Delta}_m}(S-1)$. The receive array has $N=12$ elements, which are randomly selected from the transmit array. For planar array, the reference transmit subarray is a $7 \times 7$ URA. The number of subarrays is $S = 6$, where $J = 2$ and $I = 3$. The distances between subarrays in both directions are fixed as the working wavelength, which means that $\Delta_{m_x} = 2$ and $\Delta_{m_y} = 2$. A number of $N = 12$ elements in the transmit array are randomly chosen as the receive array.

For comparisons, ESPRIT-based algorithm \cite{16} that exploits the phase rotations between transmit subarrays and U-ESPRIT algorithm \cite{19} that utilizes the conjugate symmetric property of array manifold are used as signal covariance matrix-based DOA estimation methods, while conventional ALS algorithm \cite{21,27} that decomposes the factor matrices iteratively is utilized as signal tensor decomposition-based DOA estimation method. The Cramer-Rao lower bound (CRLB) for MIMO radar is also provided. For target DOAs estimated by the factor matrix $\bf X$, if applicable, we use a postfix to distinguish it, e.g., ALS-sub (Proposed-sub) refers to the estimation result computed by $\bf X$, while ALS (Proposed) denotes the estimation result originated from $\bf G$ after tensor decomposition.

\subsection{Example 1: RMSE and Probability of Resolution for Linear Array with Non-overlapped Subarrays}
Three targets are placed at $\theta_l = [-15^\circ,5^\circ,15^\circ]$. Consider the matricized form of $\cal Z$ in \eqref{general}. The goal is to estimate $\theta_l$ from ${\bf Z} = {\bf T}_{(3)}+{\tau}{\bf R}$, where ${\bf T}_{(3)}$ is given by \eqref{T3} and ${\bf G} = {\bf K}$, the SNR is measured as: $SNR[dB] = 10\log \left( {{{\left\| {{{\bf{T}}_{(3)}}} \right\|_F^2} \mathord{\left/
 {\vphantom {{\left\| {{{\bf{T}}_{(3)}}} \right\|_F^2} {\left\| {\tau {\bf{R}}} \right\|_F^2}}} \right.\kern-\nulldelimiterspace} {\left\| {\tau {\bf{R}}} \right\|_F^2}}} \right)$. The RMSE is computed by
\begin{equation*}
    RMSE = \sqrt {\frac{1}{{2PL}}\sum\limits_{l = 1}^L {\sum\limits_{p = 1}^P {{{\left( {{{\hat \theta }_l}(p) - {\theta _l}(p)} \right)}^2}} } }.
\end{equation*}

\begin{figure}
    \centering
  \subfloat[RMSE versus SNR\label{fig1}]{%
       \includegraphics[width=0.5\linewidth]{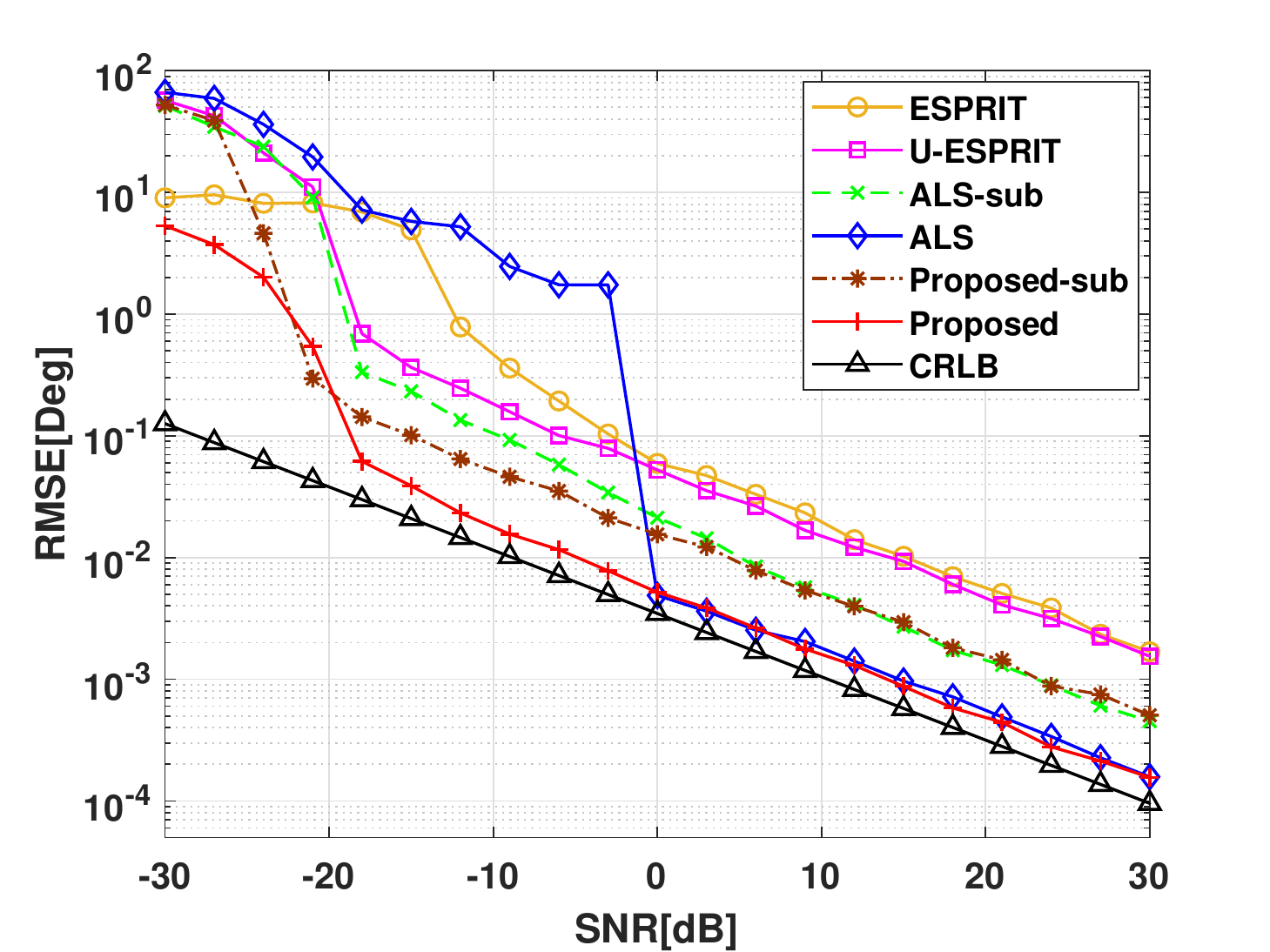}}
    \hfill
  \subfloat[Resolution versus SNR\label{fig2}]{%
        \includegraphics[width=0.5\linewidth]{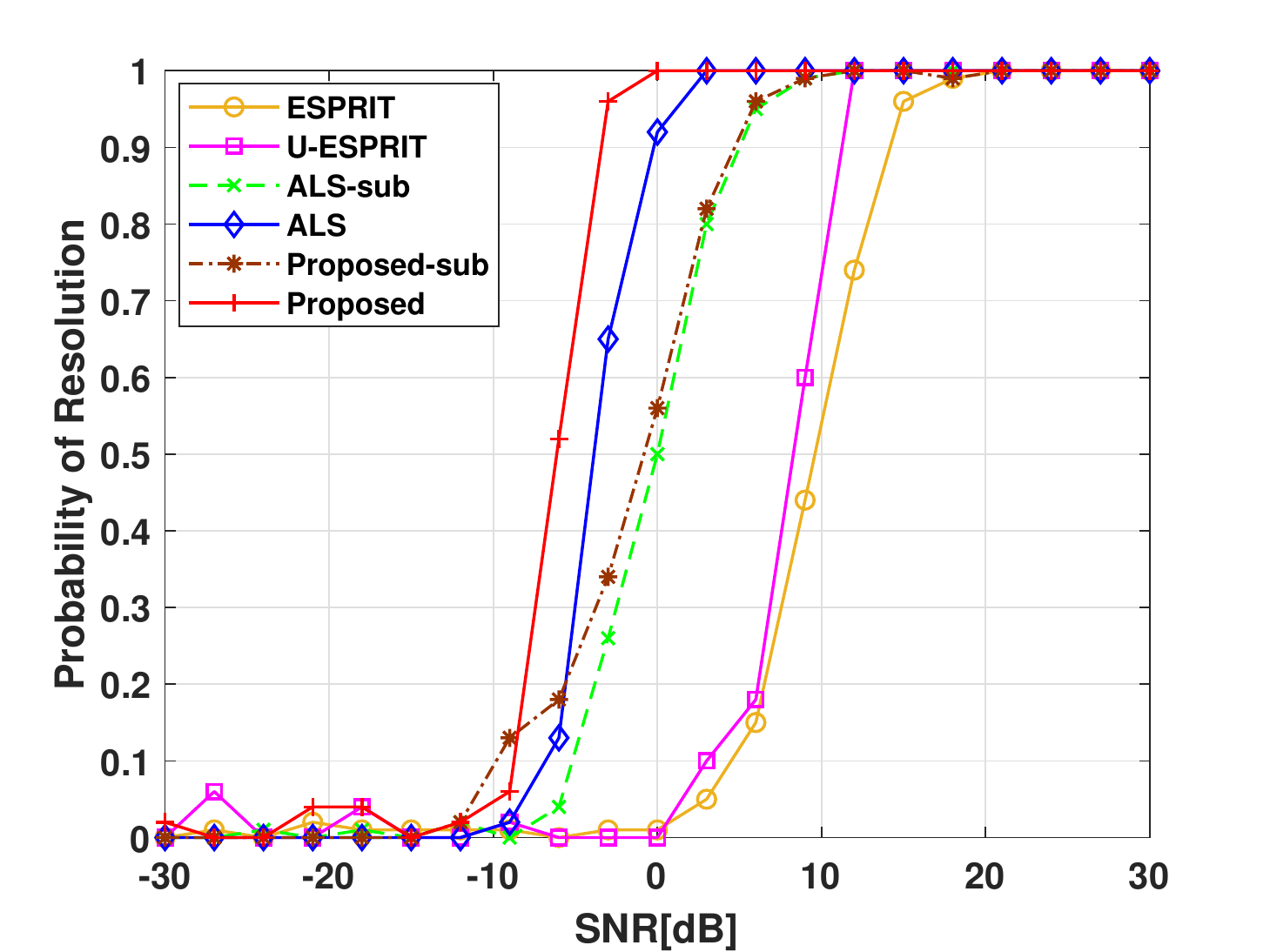}}
    \\
  \subfloat[RMSE versus SNR\label{fig5}]{%
        \includegraphics[width=0.5\linewidth]{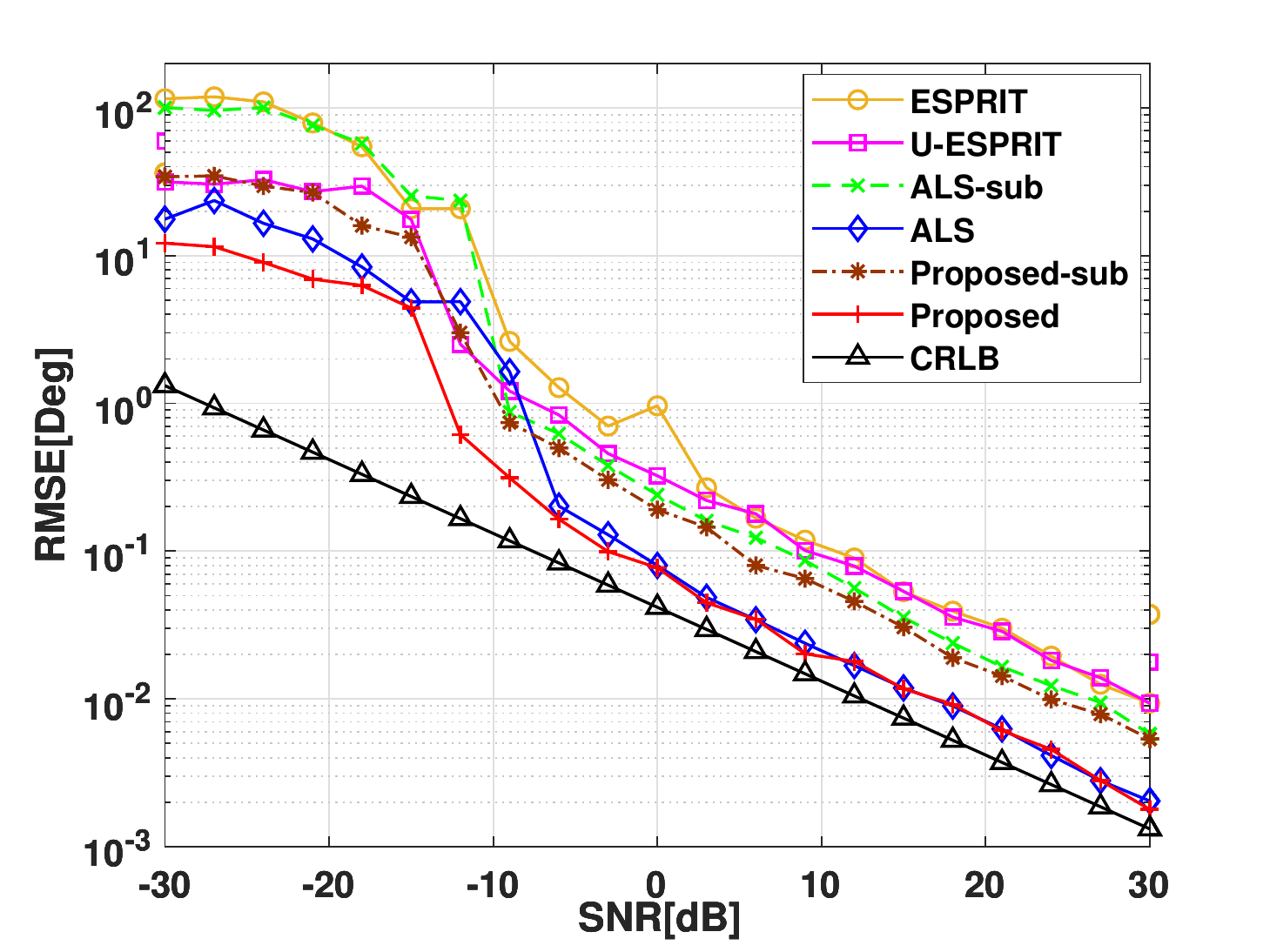}}
    \hfill
  \subfloat[Resolution versus SNR\label{fig6}]{%
        \includegraphics[width=0.5\linewidth]{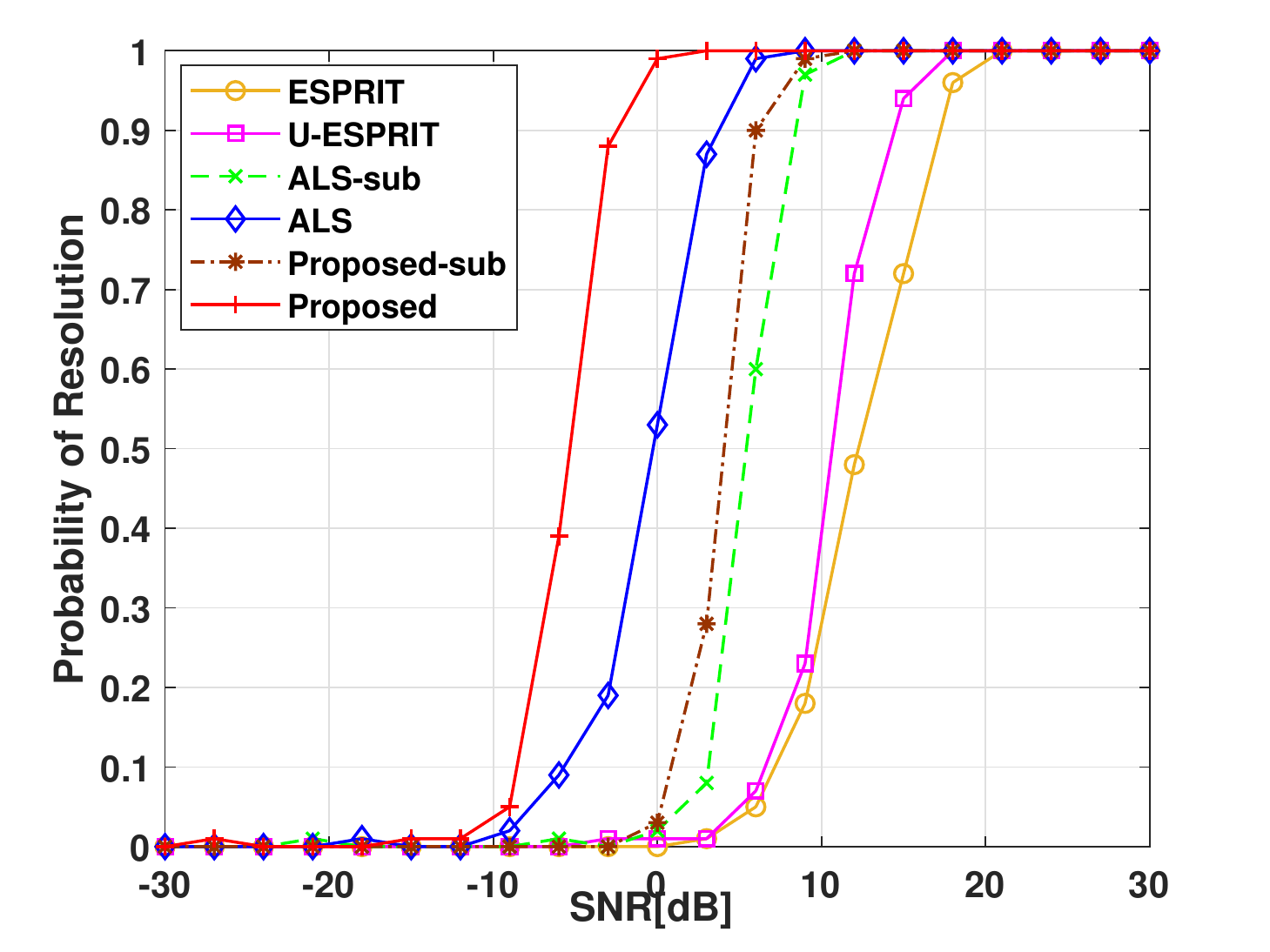}}
  \caption{DOA estimation performance for TB MIMO radar with uniformly spaced subarrays for linear array (a)-(b) and planar array (c)-(d), 200 trials.}
\end{figure}

As shown in Fig.~\ref{fig1}, the RMSE results decline gradually with the rise of SNR for all methods. The ESPRIT-based algorithm merely exploits the phase rotations between transmit subarrays and therefore the performance is quite poor. U-ESPRIT algorithm performs better since the number of snapshots is doubled. For conventional ALS algorithm and our proposed method, target angular information can be obtained from both factor matrices $\bf K$ and $\bf X$, which are used to compare to each other to eliminate the potential grating lobes. The proposed method approaches the CRLB with a lower threshold as compared to the ALS method, since the Vandermonde structure of the factor matrix is exploited. Therefore, the proposed method performs better at low SNR. Note that the complexity of our proposed method is reduced significantly, it requires approximately the same number of flops as compared to that of the ALS method in a single iteration. Also, the  comparison of the estimation results between $\bf G$ and $\bf X$ shows a reasonable difference. This is mainly caused by the different apertures of the subarray and the whole transmit array.

For the probability of resolution, we assume only two closely spaced targets located at $\theta_l = [-5^\circ,-6^\circ]$. These two targets are considered to be resolved when $\left\| {{{\hat \theta }_l} - {\theta _l}} \right\| \le \left\| {{\theta _1} - {\theta _2}} \right\|/2, l = 1,2$. The Doppler shifts are both $f = 0.2$ and the other parameters are the same as before.

In Fig.~\ref{fig2}, the probability of resolution results for all methods tested are shown and they are consistent with those in Fig.~\ref{fig1}. All methods achieve absolute resolution in high SNR region, and resolution declines with the decrease of SNR. The ESPRIT method presents the worst performance while performance of the U-ESPRIT improves slightly. The results of the ALS-sub method and the Proposed-sub method are almost the same. A gap of approximately 3~dB SNR can be observed between the proposed method and the ALS method, which means that our proposed method enables the lowest SNR threshold. The performance of both accuracy and resolution for our proposed method surpasses the other methods since the shift-invariance between and within different transmit subarrays are fully exploited.

\subsection{Example 2: RMSE and Probability of Resolution for Planar Array with Partly Overlapped Subarrays}
In this example, three targets are placed at ${\theta _l} = [ -40^{\circ}, -30^{\circ}, -20^{\circ}]$ and ${\varphi _l} = [ 25^{\circ}, 35^{\circ}, 45^{\circ}]$. The signal model ${\bf Z} = {\bf T}_{(3)}+{\tau}{\bf R}$ is applied, where ${\bf T}_{(3)}$ is given by \eqref{T3} with ${\bf G} = {\bf  H} \odot {\bf \Delta}$. The SNR is measured in the same way as that in linear array. The RMSE for planar array is compute by

\begin{footnotesize}
\begin{equation*}
    RMSE = \sqrt {\frac{1}{{2PL}}\sum\limits_{l = 1}^L {\sum\limits_{p = 1}^P {\left[ {{{\left( {{{\hat \theta }_l}(p) - {\theta _l}(p)} \right)}^2} + {{\left( {{{\hat \varphi }_l}(p) - {\varphi _l}(p)} \right)}^2}} \right]} } }.
\end{equation*}
\end{footnotesize}

In Fig.~\ref{fig5}, the RMSEs of ESPRIT, U-ESPRIT, ALS and the proposed method are given. The CRLB is also provided. The performance of the ESPRIT method and the U-ESPRIT method are relatively poor. It is because of the ignorance of the received signal shift-invariance within a single subarray. It can be observed that the Proposed-sub and the ALS-sub successfully estimate the target DOAs via $\bf X$ in the case of planar array, which proves the validity of  \eqref{convex}. The results can be used to mitigate the spatial ambiguity in the following estimations. Like their counterparts in linear array, the RMSEs of the proposed method and the ALS method are almost the same for above 0~dB SNR while the performance of the proposed method in low SNR is better. The ALS method ignores the Vandermonde structure during tensor decomposition. Compared to \eqref{convex}, the DOA estimation result in ${\bf G}$ takes advantage of a larger aperture and therefore achieves a better RMSE performance.

To evaluate the resolution performance, only two targets are reserved and the spatial directions are ${\theta _l} = [ -10^{\circ}, -11^{\circ}]$ and ${\varphi _l} = [ 15^{\circ}, 16^{\circ}]$. The resolution is considered successful if $\left\| {{{\hat \theta }_l} - {\theta _l}} \right\| \le \left\| {{\theta _1} - {\theta _2}} \right\|/2, \left\| {{{\hat \varphi }_l} - {\varphi _l}} \right\| \le \left\| {{\varphi _1} - {\varphi _2}} \right\|/2, l = 1,2$. The target Doppler shifts are the same, given as $f = 0.2$. The other parameters are unchanged.

Fig.~\ref{fig6} shows the results for all methods with respect to the probability of resolution. The proposed method achieves the lowest SNR threshold, which benefits from the fully exploitation of the shift-invariance and the Vandermonde structure during tensor decomposition. Note that the convergence of the ALS method is unstable and can be influenced by the tensor size. It can be observed that the resolution performance of the ALS method is deteriorated as compared to its counterpart in Fig.~\ref{fig2}. This conclusion implies that the robustness of our proposed method is better regarding 2-D DOA estimation, since no iterations are required.

\subsection{Example 3: RMSE Performance for Linear Array with Different ${\Delta}_m $}
In this example, we mainly consider the RMSE performance when $\Delta_m$ changes from one to at most $M_0$. The aperture is increased gradually. The SNR is assumed to be 10~dB. All other parameters are the same as those in Example~1.

Given the number of subarrays and the structure of a single subarray, the aperture of the overall transmit ULA rises with the increase of $\Delta_m$ while the number of elements shared by two adjacent subarrays declines. When $\Delta_m = 0$, this model is identical to that for conventional ESPRIT method \cite{12,31}, and there is no transmit subarray. When $\Delta_m$ rises, the distance between phase centers for two adjacent subarrays becomes larger than half the working wavelength and grating lobes are generated. The locations of these grating lobes are determined by \eqref{grating}, and can be eliminated. Meanwhile, the transmit array aperture is increased and the DOA estimation performance should be improved.

\begin{figure}
    \centering
  \subfloat[RMSE versus ${\Delta}_m$\label{fig3}]{%
       \includegraphics[width=0.5\linewidth]{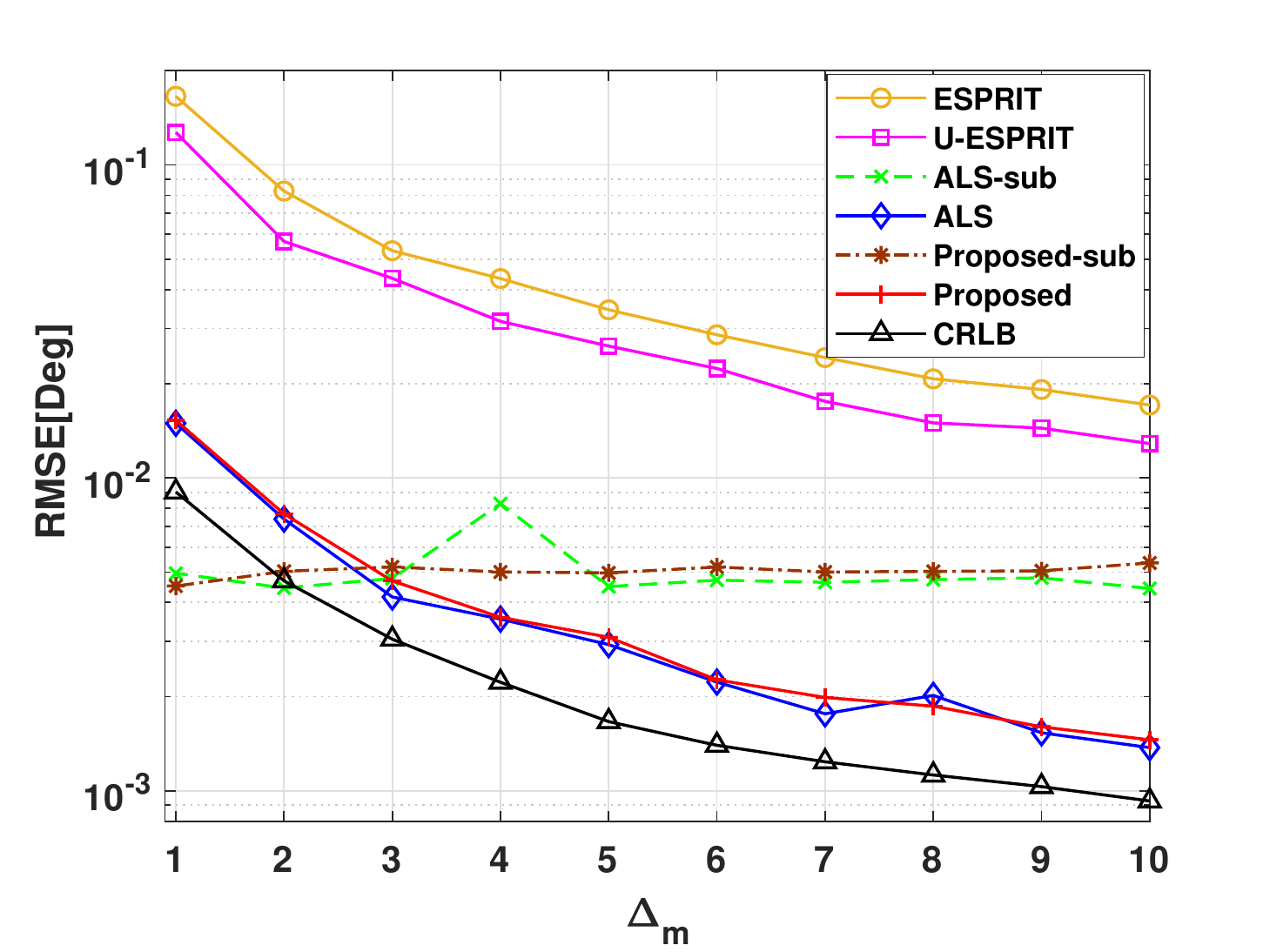}}
    \hfill
  \subfloat[Generalized Vandermonde matrix\label{fig4}]{%
        \includegraphics[width=0.5\linewidth]{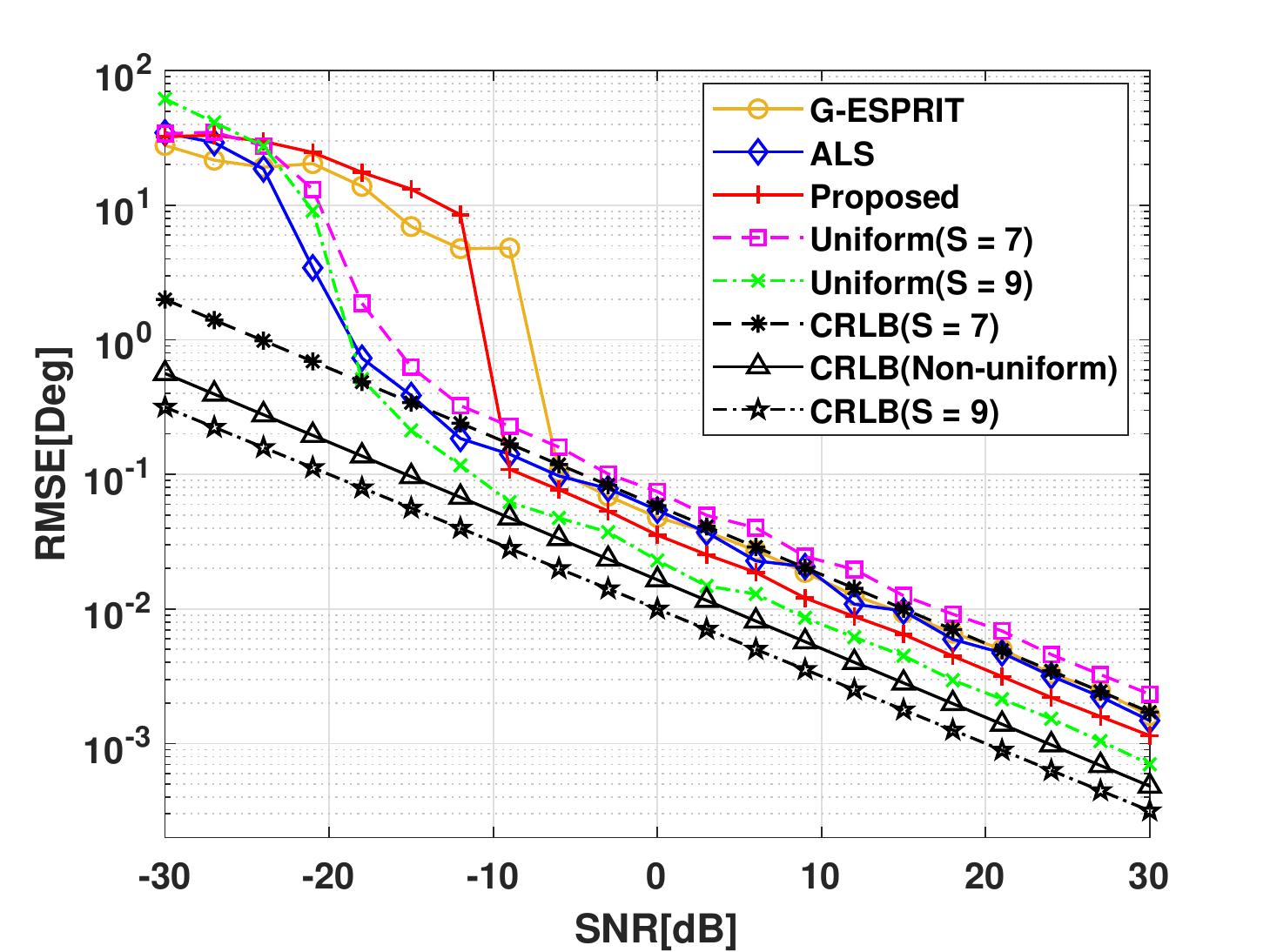}}
    \\
  \subfloat[Elevation RMSE versus SNR\label{fig7}]{%
        \includegraphics[width=0.5\linewidth]{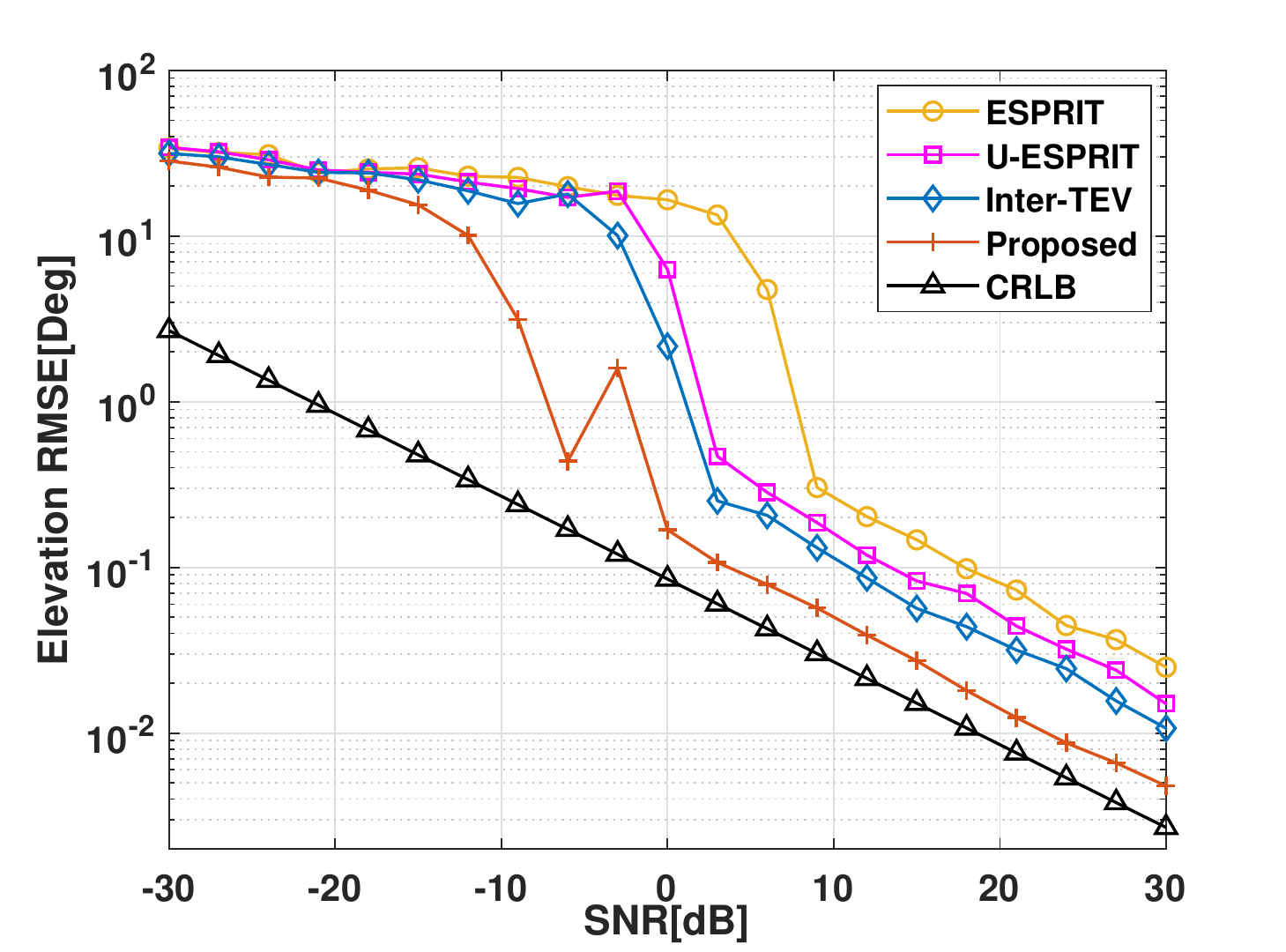}}
    \hfill
  \subfloat[Azimuth RMSE versus SNR\label{fig8}]{%
        \includegraphics[width=0.5\linewidth]{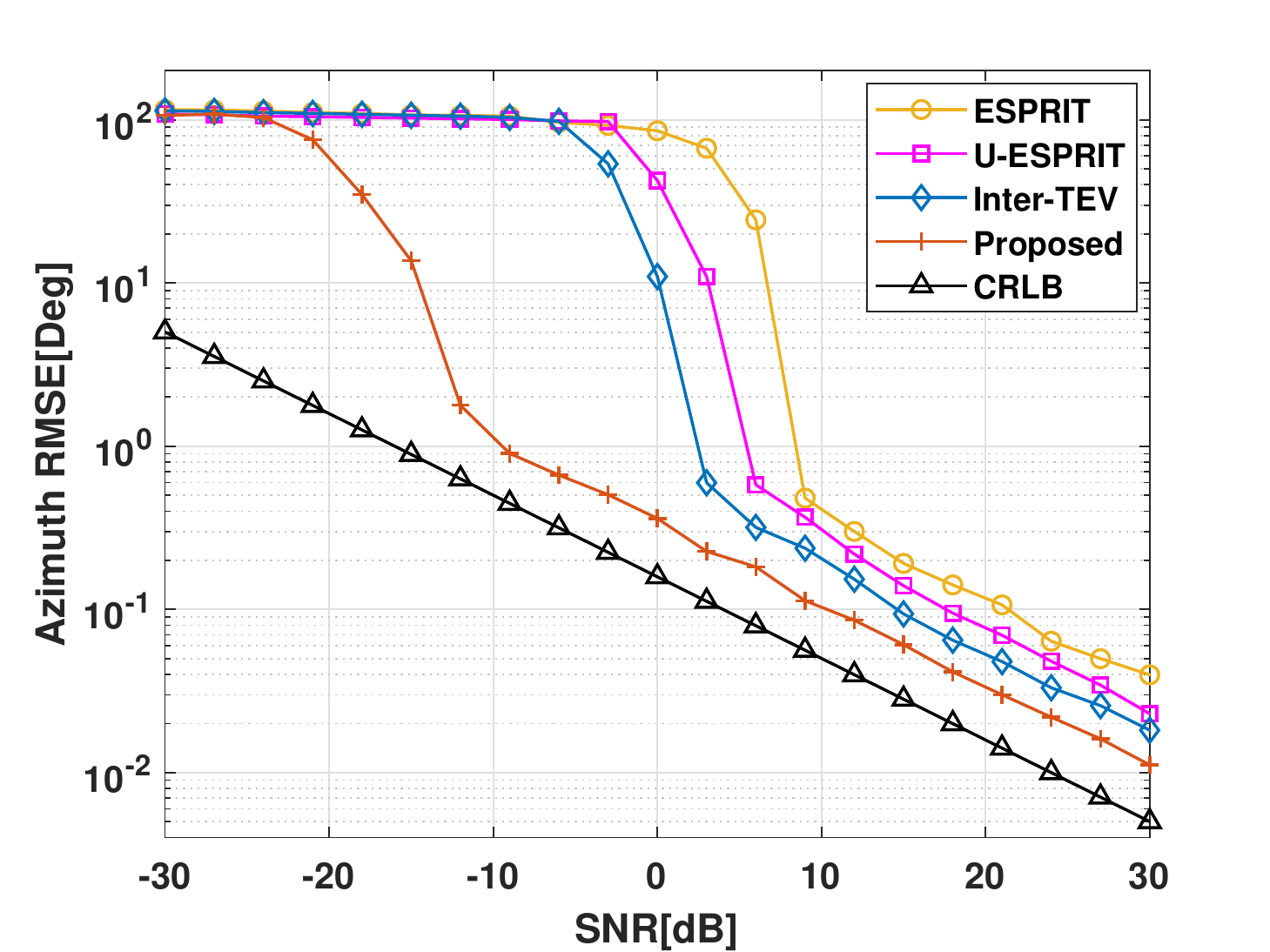}}
  \caption{ RMSE results for TB MIMO radar with different subarray configurations.}
\end{figure}

To investigate the improvement, the RMSEs of three targets are computed versus the rise of $\Delta_m$. It can be seen in Fig.~\ref{fig3} that the RMSE results decrease steadily with the increase of $\Delta_m$. The ESPRIT method and U-ESPRIT method suffer from grating lobes and the received signal within a single subarray is not fully exploited, hence, they perform poorly. The RMSEs of the Proposed-sub and the ALS-sub are almost unchanged since the estimation is only based on the subarray, which is fixed during the simulation. Meanwhile, the proposed method and the ALS method achieve better accuracy than their counterparts originated from $\bf X$ when $\Delta_m >3$. It can be noted in Fig.~\ref{fig1} that the convergence is satisfied for the ALS method and our proposed method when SNR is above 10~dB. Consequently, the RMSEs of the proposed method and the ALS method are nearly coincident.

To evaluate the RMSE performance versus $\Delta_{m_x}$ or $\Delta_{m_y}$ for a planar array, it is necessary to separately add a new subarray in one direction while keeping the array structure in the other direction unchanged. This can be fulfilled by constructing an L-shaped transmit array, where each element is replaced by a URA subarray. However, this analysis would be beyond the scope of this paper. In general, it can be concluded that the proposed method can estimate the target DOAs via the phase rotations between transmit subarrays. If the placement of two adjacent subarrays satisfies some conditions, e.g., $\Delta_m >3$ for linear array, the RMSE performance is better than that computed by a single subarray. Note that the received signal of two adjacent subarrays can be obtained by spatial smoothing \cite{32}, a proper spatial smoothing of the received signal can improve the DOA estimation performance.

\subsection{Example 4: Generalized Vandermonde Factor Matrix for Linear Array with $m_s = \{1,2,3,5,7,9\}$}
Here we evaluate the proposed DOA estimation method for TB MIMO radar with non-uniformly spaced transmit subarrays. The transmit linear array has $S = 7$ subarrays with $m_s = \{1,2,3,5,6,7,9\}$. Each subarray contains $M_0 = 10$ elements. The $N = 12$ elements are randomly chosen from the transmit array to form the receive array. Three targets are placed at ${\theta_l} = [-5^\circ,10^\circ,18^\circ]$ with normalized Doppler shifts $f_l = [0.3,-0.15,-0.15]$. To simplify the signal model, each subarray is a ULA, which is not used during the DOA estimation in this example. Equations \eqref{case}-\eqref{43} can be applied directly, since the subarray structure stays identical. Two different transmit arrays are introduced for comparison to illustrate the improved performance provided by constructing ${\bf K}^{(sub)}$. Both of them can be regarded as a linear array with uniformly spaced subarrays ($\Delta_m = 1$). The first one has $S = 7$ subarrays, while the second one has $S = 9$ subarrays to achieve the same aperture. The DOA estimation for these two transmit arrays can be conducted by \textbf{Algorithm~\ref{alg1}}. Meanwhile, conventional ALS method can be applied to decompose the factor matrix ${\bf K}^{(sub)}$, which will be used to estimate the target DOAs by solving \eqref{43}. The generalized-ESPRIT (G-ESPRIT) method in \cite{46} is also used for comparison. The CRLBs of three different transmit arrays are also shown.

From Fig~\ref{fig4}, it can be observed that the formulation of ${\bf K}^{(sub)}$ exploits the multiple scales of shift-invariances in generalized Vandermonde matrix. By solving \eqref{43}, the grating lobes are eliminated efficiently. Hence, the structurs of transmit subarrays can be arbitrary but identical, which provide more flexibility for array design. The RMSE of the proposed method surpasses those of G-ESPRIT and ALS methods. Also, the performance of the non-uniformly spaced transmit subarrays is better than that of the uniformly spaced transmit subarrays (S = 7). This is expected since the aperture is increased due to sparsity. Compared to the fully spaced transmit subarray case (S = 9), the performance of the proposed method is deteriorated slightly. However, the fully spaced array can be extremely high-cost if the array aperture is further increased. By using the generalized Vandermonde matrix, the proposed method enables the sparsity in transmit array, which achieves higher resolution with less elements.

\subsection{Example 5: Multiscale Sensor Array with Arbitrary but Identical Subarrays}
In the final example, we illustrate the performance of the proposed DOA estimation method for TB MIMO radar with arbitrary but identical subarrays. Specifically, a planar array with $S = 4\times 4$ subarrays is considered, whose phase centers form a uniform rectangular grid with a distance of half the working wavelength. For each subarray, $M_0 = 4$ elements are randomly placed in a circle centered on the phase center with a radius of a quarter of wavelength. The structure of all subarrays are identical, hence, the transmit array can be regarded as an 3-level multiscale sensor array and the transmit subarray steering matrix can be obtained from \eqref{G}. The $N  =12 $ receive elements are randomly selected from the transmit array. Three targets are placed at ${\theta _l} = [ -26^{\circ}, -19^{\circ}, -12^{\circ}]$ and ${\varphi _l} = [ 11^{\circ}, 21^{\circ}, 31^{\circ}]$. The other parameters are the same as those in Example~2.

Note that the subarray is arbitrary, the DOA information can only be estimated by the phase rotations between the transmit subarrays. Alternatively, the transmit array interpolation technique \cite{23} is introduced to map the original transmit array into a $4 \times 4$ URA to enable the ESPRIT-like DOA estimation, which is referred to as Inter-TEV in Fig.~\ref{fig7} and Fig.~\ref{fig8}. It can be observed that by carefully designing the mapping matrix, the RMSEs of the Inter-TEV method are better than those of ESPRIT and U-ESPRIT methods for both elevation and azimuth estimation. However, the proposed method surpasses the other methods with a lower RMSE. This is because of the full usage of the shift-invariance between and within different transmit subarrays.

\section{Conclusion}\label{6}
The problem of tensor decomposition with Vandermonde factor matrix in application to DOA estimation for TB MIMO radar with arbitrary but identical transmit subarrays has been considered. A general 4-order tensor that can be used to express the TB MIMO radar received signal in a variety of scenarios, e.g., linear and planar arrays, uniformly and non-uniformly spaced subarrays, regular and irregular subarrays, has been designed. The shift-invariance of the received signal between and within different transmit subarrays have been used to conduct DOA estimation. Specifically, a computationally efficient tensor decomposition method has been proposed to estimate the generators of the Vandermonde factor matrices, which can be used as a look-up table for finding target DOA. The proposed method fully exploits the shift-invariance of the received signal between and within different subarrays, which can be regarded as a generalized ESPRIT method. Comparing with conventional signal tensor decomposition-based techniques, our proposed method take advantage of the Vandermonde structure of factor matrices, and it requires no iterations or any prior information about the tensor rank. The parameter identifiability of our tensor model has also been studied via the discussion of the uniqueness condition of tensor decomposition. Simulation results have verified that the proposed DOA estimation method has better accuracy and higher resolution as compared to existing techniques.

\appendices
\setcounter{section}{0}
\setcounter{equation}{45}
\vspace{-3mm}

\section{Proof of \textbf{Lemma~\ref{111}}}\label{222}
First, let ${\bf A}^{(1)} \in {\mathbb{C}^{{I_1}\times L}} $ be a Vandermonde matrix with distinct generators, we have $r\left({\bf A}^{(1)} \odot {\bf A}^{(2)}\right) = \min({I_1I_2,L})$, since it is the KR product of a Vandermonde matrix and an arbitrary matrix~[17].  Assuming $I_3 \geq L, I_1I_2 \geq L$, the following results $r\left({\bf A}^{(1)} \odot {\bf A}^{(2)}\right) = L$ and $r({\bf A}^{(3)}) = L$ hold, since ${\bf A}^{(3)}$ has full column rank. The proof of \textbf{Lemma~\ref{111}} in this case is identical to that of Proposition~III.2 in~[17].

Next, let ${\bf A}^{(1)} = {\bf B} \odot {\bf C}$, where ${\bf B} \in {\mathbb{C}^{{J}\times L}}$ and ${\bf C} \in {\mathbb{C}^{{I}\times L}}$ are both Vandermonde matrices with distinct generators. Consider the rank of matrix ${\bf A}^{(1)} \odot {\bf A}^{(2)} = {\bf B} \odot {\bf C} \odot {\bf A}^{(2)} = {\bf \Pi}\left({\bf B} \odot {\bf A}^{(2)} \odot {\bf C}\right)$, where ${\bf \Pi}$ is an exchange matrix. Again, the rank of ${\bf B} \odot {\bf A}^{(2)}$ is $\min (JI_2, L)$ while $r\left({\bf B} \odot {\bf A}^{(2)} \odot {\bf C}\right) = \min (IJI_2, L)$.  Since ${\bf \Pi}$ is nonsingular, $r\left({\bf A}^{(1)} \odot {\bf A}^{(2)}\right) = L$.

The mode-3 unfolding of $\cal Y$ is ${\bf Y}_{(3)} = \left({\bf A}^{(1)} \odot {\bf A}^{(2)}\right){\bf A}^{(3)T}$. The SVD of this matrix representation is denoted by ${\bf Y}_{(3)} = {\bf U}{\bf \Lambda}{\bf V}^H$, where ${\bf{U}} \in {\mathbb{C}^{{I_1I_2} \times L}}$, ${\bf{\Lambda}} \in {\mathbb{C}^{L \times L}}$, and ${\bf{V}} \in {\mathbb{C}^{{I_3} \times L}}$. Since $r\left({\bf A}^{(1)} \odot {\bf A}^{(2)}\right) = L$ and $r\left({\bf A}^{(3)}\right) = L$, it can be derived that a nonsingular matrix ${\bf E}\in {\mathbb{C}^{{L}\times L}}$ satisfies
\begin{equation}
{\bf U}{\bf E} = {\bf A}^{(1)} \odot {\bf A}^{(2)}.
\end{equation}

The Vandermonde structure of both ${\bf B}$ and ${\bf C}$ can be exploited via
\begin{equation}
\begin{aligned}
& {{\bf{U}}_{\rm{2}}}{\bf{E}} = {\bf{\underline B }} \odot {\bf{C}} \odot {\bf A}^{(2)} = \left( {{\bf{\overline B }} \odot {\bf{C}} \odot {\bf A}^{(2)}} \right){{\bf{\Omega }}_b} = {{\bf{U}}_1}{\bf{E}}{{\bf{\Omega }}_b}\\
& {{\bf{U}}_{\rm{4}}}{\bf{E}} = {\bf{B }} \odot {\bf{\underline C}} \odot {\bf A}^{(2)} = \left( {{\bf{B}} \odot {\bf{\overline C}} \odot {\bf A}^{(2)}} \right){{\bf{\Omega }}_c} = {{\bf{U}}_3}{\bf{E}}{{\bf{\Omega }}_c}\\
\end{aligned}
\end{equation}
where ${{\bf{\Omega }}_b} = diag({\bm \omega}_b)$, ${{\bf{\Omega }}_c} = diag({\bm \omega}_c)$ with ${\bm \omega}_b$ and ${\bm \omega}_c$ denoting the vectors of generators of $\bf B$ and $\bf C$, respectively. The submatrices ${{\bf{U}}_{\rm{1}}}$, ${{\bf{U}}_{\rm{2}}}$, ${{\bf{U}}_{\rm{3}}}$ and ${{\bf{U}}_{\rm{4}}}$ are truncated from rows of ${\bf{U}}$ according to the operator of the KR product, i.e.,
\begin{equation}
\begin{aligned}
&{{\bf{U}}_1}{\rm{ = }}\left[ {{{\bf{I}}_{{I}K({J} - 1)}},{{\bf{0}}_{{I}K({J} - 1) \times {I}K}}} \right]{\bf{U}}\\
&{{\bf{U}}_2}{\rm{ = }}\left[ {{{\bf{0}}_{{I}K({J} - 1) \times {I}K}},{{\bf{I}}_{{I}K({J} - 1)}}} \right]{\bf{U}}\\
&{{\bf{U}}_3} = \left( {{{\bf{I}}_{{J}}} \otimes \left[ {{{\bf{I}}_{K({I} - 1)}},{{\bf{0}}_{K({I} - 1) \times K}}} \right]} \right){\bf{U}}\\
&{{\bf{U}}_4} = \left( {{{\bf{I}}_{{J}}} \otimes \left[ {{{\bf{0}}_{K({I} - 1) \times K}},{{\bf{I}}_{K({I} - 1)}}} \right]} \right){\bf{U}}.\\
\end{aligned}
\end{equation}

Note that $\bf E$, ${\bf \Omega}_b$ and ${\bf \Omega}_c$ are full rank. We have ${\bf{U}}_1^\dag {{\bf{U}}_2} = {\bf{E}}{{\bf{\Omega }}_b}{\bf E}^{-1}$ and ${\bf{U}}_3^\dag {{\bf{U}}_4} = {\bf{E}}{{\bf{\Omega }}_c}{\bf E}^{-1}$. Hence, the vectors ${\bm \omega}_b$ and ${\bm \omega}_c$ can be computed as the collections of eigenvalues of ${\bf{U}}_1^\dag {{\bf{U}}_2}$ and ${\bf{U}}_3^\dag {{\bf{U}}_4}$, respectively, while $\bf E$ is the matrix of collection of the corresponding eigenvectors. From the generators of $\bf B$ and $\bf C$, the first factor matrix ${\bf A}^{(1)}$ can be reconstructed.

Meanwhile, it can be observed that
\begin{equation}
  \left( {\frac{{{\bm{\alpha }}_l^{(1)H}}}{{{\bm{\alpha }}_l^{(1)H}{{\bm{\alpha }}_l^{(1)}}}} \otimes {{\bf{I}}_{I_2}}} \right)\left( {{{\bm{\alpha }}_l^{(1)}} \otimes {{\bm{\alpha }}_l^{(2)}}} \right) = {{\bm{\alpha }}_l^{(2)}}.
\end{equation}

Assume that the column vectors of ${\bf A}^{(1)}$ have unit form, then ${\bm \alpha}_l^{(2)}$ can be written as
\begin{equation}
{\bm \alpha}_l^{(2)} = ({\bm \alpha}_l^{(1)H} \otimes {\bf I}_{I_2}){\bf U}{\bf e}_l,\quad l = 1,2,\cdots, L
\end{equation}
where ${\bf e}_l$ is the $l$-th column of $\bf E$. Given ${\bf A}^{(1)}$ and ${\bf A}^{(2)}$, the third factor matrix can be computed by
\begin{equation}
\begin{aligned}
{\bf A}^{(3)T} = {\left( {\left( {{{\bf{A}}^{(1)H}}{{\bf{A}}^{(1)}}} \right) * \left( {{{\bf{A}}^{(2)H}}{{\bf{A}}^{(2)}}} \right)} \right)^{ - 1}} \\
\times {\left( {{{\bf{A}}^{(1)}} \odot {{\bf{A}}^{(2)}}} \right)^H}{{\bf{Y}}_{(3)}}.
\end{aligned}
\end{equation}

Therefore, the tensor decomposition of $\cal Y$ is generically unique, where ${\bf A}^{(1)}$ can be a Vandermonde matrix or the KR product of two Vandermonde matrices with distinct generators, and ${\bf A}^{(3)}$ is column full rank.

\section{Proof of \eqref{2Dunfolding}}\label{A}
Given ${{\bf{y}}^{(q)}_{s}}$ in \eqref{z}, $s = 1,2,\cdots, I,I+1, \cdots, IJ$, concatenate every $I$ vectors together to form totally $J$ matrices of identical dimension $KN \times I$. These matrices are denoted by ${\bf \bar Y}^{(q)}_{j}$, and are given as
\begin{equation}
{\bf \bar Y}^{(q)}_{j} = \left[ {\left( {{\bf{W}}_0^H{{\bf{A}}_0}} \right) \odot {\bf{B}}} \right]{\left( {{\bf{c}}_q^T \odot {\bf{h}}^T_{{j}} \odot {\bf{\Delta}}} \right)^T} + {{\bf{\bar N}}^{(q)}_{j}}\label{xy}
\end{equation}
where ${{\bf{\bar N}}_j^{(q)}} \triangleq \left[ {{{\bf{n}}_{(j-1)I+1}^{(q)}},{{\bf{n}}_{(j-1)I+2}^{(q)}},\cdots,{{\bf{n}}_{(jI)}^{(q)}}} \right]$. The noise-free version of \eqref{xy} can be rewritten as
\begin{equation}
{\bf \bar Y}^{(q)}_{j}  = \left[ {\left( {{\bf{W}}_0^H{{\bf{A}}_0}} \right) \odot {\bf{B}}} \right]{\bf{\Gamma }}_q{\left( { {{\bf{h}}^T_{{j}}} \odot {\bf{\Delta }} } \right)^T}
\end{equation}
where ${\bf{\Gamma }}_q = diag({\bf c}_q)$. Since $\left[ {\left( {{\bf{W}}_0^H{{\bf{A}}_0}} \right) \odot {\bf{B}}} \right]{\bf{\Gamma }}_q$ is fixed, the concatenation of $J$ matrices merely depends on the concatenation of ${\left( { {{\bf{h}}^T_{{j}}} \odot {\bf{\Delta }} } \right)^T}$. Define a matrix ${\bf \Theta}$
\begin{equation}
{\bf{\Theta }} \triangleq {\left[ {\begin{array}{*{20}{c}}
{{\bf{h}}_1^T \odot {\bf{\Delta }}}\\
{{\bf{h}}_2^T \odot {\bf{\Delta }}}\\
 \vdots \\
{{\bf{h}}_J^T \odot {\bf{\Delta }}}
\end{array}} \right]_{S \times L}}
\end{equation}
 such that ${\bf{\Theta }}$ performs the concatenation. From the definition of the KR product, it can be observed that ${\bf{\Theta }} = {\bf H} \odot {\bf \Delta}$. Therefore, the concatenation of ${\bf \bar Y}^{(q)}_{j}$ is given by
\begin{equation}
\begin{aligned}
{\bf \bar Y}^{(q)} &= \left[ {\left( {{\bf{W}}_0^H{{\bf{A}}_0}} \right) \odot {\bf{B}}} \right]{\bf{\Gamma }}_q{\bf{\Theta }}^T\\
& = \left[ {\left( {{\bf{W}}_0^H{{\bf{A}}_0}} \right) \odot {\bf{B}}} \right]{\left( {{{\bf{c}}_q^T} \odot {\bf{H }} \odot {\bf{\Delta}}} \right)^T}.
\end{aligned}
\end{equation}

Considering the noise term, equation \eqref{2Dunfolding} can be given.

\bibliographystyle{IEEEtran}
\bibliography{IEEEabrv,Ref}

\end{document}